\documentclass[useAMS,usenatbib,]{mn2e}
\usepackage{graphicx}
\usepackage{txfonts}
\usepackage{natbib}
\usepackage{hyperref}
\usepackage{times,amssymb,amsfonts,amsbsy}

\begin{document}

\title[On the concentration-mass relation]
  {The effects of baryonic cooling on the concentration-mass relation}

\author[C. Fedeli]{C. Fedeli\\
	Department of Astronomy, University of Florida, 211 Bryant Space Science Center, Gainesville, FL 32611 \emph{(cosimo.fedeli@astro.ufl.edu)}}

\maketitle

\begin{abstract}
I re-examine the relation between virial mass and concentration for groups and clusters of galaxies as measured in a number of recent works. As previously noted by several authors, low-mass clusters and groups of galaxies display systematically larger concentrations than simple prescriptions based on pure $n$-body simulations would predict. This implies an observed concentration-mass relation with a substantially larger slope/normalization than expected from theoretical investigations. Additionally, this conclusion seems to be quite independent on selection effects, holding for both lensing based and X-ray based cluster samples. In order to shed new light on this issue I employ a simple spherical halo model containing, in addition to dark matter, also stars and hot diffuse gas in proportions and with distributions in agreement with the most recent observations. Moreover, I include the contraction effect experienced by dark matter due to the cooling of baryons in the very central part of the structure itself. The resulting modified concentration-mass relation is steeper than the theoretical input one, because star formation is fractionally more efficient in low-mass objects. However, the effect is non-vanishing at all masses, thus resulting also in a larger normalization. Overall the new relation provides a better representation of the observed one for almost all catalogs considered in this work, although the specific details depend quite significantly on the baryon fraction prescription adopted. Specifically, the observed concentration-mass relation seems to favor a scenario where the stellar mass fraction in large clusters of galaxies is substantially lower than several works have found. Anyhow, the same effect could also be produced by a redistribution of baryons within the structure. Moreover, the concentration of a number of high-mass objects seems to be significantly lower even than the predictions based on pure $n$-body simulations, and they are hence unaccounted for in the modified scenario that is proposed here. Finally I use this simple model to show how the estimated concentration of cosmic structures is expected to be overestimated as a function of the radial range covered by the analysis.
\end{abstract}

\section{Introduction}\label{sct:introduction}

One of the outstanding successes of the standard $\Lambda$CDM cosmological model \citep{RU09.1,KO11.1} is its account for the formation of structures in the Universe. Accordingly, tiny density fluctuations were laid down in the dark matter distribution during the early stages of the Universe \citep{HA83.1}. These seed perturbations grew up in time due to gravitational instability, and eventually detached away from the overall expansion and collapsed, giving rise to the web of clumps, filaments, and voids that compose the currently observed large scale structure (e.g., \citealt{PE01.1,PE07.1}).

The redshift evolution of dark matter structures is determined only by gravity, and it is hence quite easy to model with the help of numerical $n$-body simulations. As a matter of fact, simulations provide a plethora of predictions that have been tested against observations, up to now with an overall good degree of success. One of the principal predictions is that dark matter bound structures (or \emph{halos}) should have a cuspy density profile, on average well represented by the two-parameter \citet*{NA96.1} (NFW henceforth, see also \citealt*{DU91.1,NA95.1,NA97.1}) function

\begin{equation}
\rho(x) = \frac{\rho_\mathrm{s}}{x~(1+x)^2}~,
\end{equation}
where $\rho_\mathrm{s}$ is a scale density, while $x\equiv r/r_\mathrm{s}$ is the radial distance from the center of the halo in units of a scale radius $r_\mathrm{s}$. The two parameters of the NFW profile can be easily related to more directly measurable quantities, namely the mass and the \emph{concentration} of the halo. The mass of a dark matter halo is conventionally defined as the mass contained within a sphere whose average density equals some factor $\Delta$ times the critical density of the Universe at the redshift of interest,

\begin{equation}
m_\Delta = \frac{4}{3}\pi R_\Delta^3 \left[\Delta\rho_\mathrm{c}(z)\right]~.
\end{equation}
The specific value of $\Delta$ varies according to the author. The most common approach is to set $\Delta = 200$, independently on redshift and cosmology. It has been shown that the $R_{200}$ thus defined is a good approximation of the radius within witch the structure is in dynamical equilibrium \citep*{EK98.1}. Another popular approach is to have $\Delta = \Delta_\mathrm{v}(z)$, where $\Delta_\mathrm{v}(z)$ is the virial overdensity according to the spherical collapse model, which depends on redshift and cosmology \citep{BR98.1}. Also, a number of authors prefer to replace the critical density of the Universe with the mean matter density. In this work I adopted the former choice, with $\Delta=200$. The concentration of a halo is a measure of its compactness, and is defined for a NFW profile as $c_\Delta\equiv R_\Delta/r_\mathrm{s}$. In what follows I will drop the suffix $\Delta$ in all quantities, leaving implicit that $\Delta = 200$, and I shall refer to $m=m_{200}$ and $R=R_{200}$ as to \emph{virial} quantities.

Another cornerstone prediction of the standard cosmology is the hierarchical clustering, meaning the fact that high-mass objects are formed as aggregates of lower-mass ones, that were formed at earlier times in the same manner. A related feature is that the concentration and the mass of a dark matter structure are not unrelated, rather the concentration is a decreasing function of mass. This result is commonly interpreted as due to the fact that halos retain information on the average matter density of the Universe at the moment of their formation \citep*{NA96.1}. Since low-mass structures form earlier than large-mass ones, they have more time to relax and compactify, hence displaying a larger concentration. The concentration-mass relation predicted by $n$-body simulations is well represented by a shallow power law $c (m,z=0) \propto m^{-\alpha}$, with $\alpha \sim 0.1$ \citep{DO04.1,GA08.1,ZH09.2}. Moreover, the distribution around this mean relation is basically log-normal, with a logarithmic mean deviation ranging from $\sim 0.15$ for the most relaxed structures, up to $\sim 0.30$ for the most disturbed ones \citep{JI00.1}.

The redshift dependence of the concentration for a given mass is much less established. While the original NFW prescription displays very little redshift evolution, simple arguments based on later studies (\citealt{BU01.1}; \citealt*{EK01.1}) suggest that the concentration of a dark matter halo with fixed mass should evolve quite strongly with redshift, basically $\propto (1+z)^{-1}$. More recent studies however \citep{NE07.1,DU08.1,GA08.1,ZH09.2,PR11.1} imply that the original NFW prediction was not so wrong after all. The concentration of high-mass objects shows basically no evolution with redshift, while the overall normalization evolves much less strongly than $\propto (1+z)^{-1}$, as previously thought.

Substantial complication is added to this picture by the role of baryonic matter, that is gas and stars. Although these luminous matter components are almost always subdominant with respect to dark matter, and hence are not expected to drive the process of structure formation, they are well known to have an impact on the details of matter allotment (\citealt{JI06.1}; \citealt*{FE11.1}). While on very large scales (typically above a few tens of Mpc) it is safe to assume that the gas distribution follows the dark matter one, on small scales (say below $0.1$ Mpc) the gas becomes dense enough to cool down radiatively and condense into stars. Massive stars themselves subsequently provide energy and metal injection into the gas when they go off as supernovae, with Active Galactic Nuclei (AGN) activity constituting another important source of energetic feedback. All these non-gravitational processes are very difficult to model, and the majority of them takes place at scales that are too small to be resolved by current cosmological simulations, thus requiring the implementation of sub-grid approximations. 

As of today, no consensus has been reached as for the impact of baryons on the small scale matter distribution. Yet, observations are now achieving a precision level where such an impact cannot be neglected anymore. Examples in this direction are given by future wide field weak lensing surveys (such as \emph{Euclid} and WFIRST, \citealt{LA11.1}) that will be capable of measuring the non-linear matter power spectrum at the percent level, or the upcoming HST CLASH program \citep{PO11.1}, that will determine the internal structure of massive galaxy clusters with a relative accuracy of $\sim 5\%$. In both cases the impact of baryons is expected to be substantially larger than the forecasted measurement precisions.

In this paper I used a spherical semi-analytic model in order to investigate the impact of baryonic physics, specifically gas cooling and star formation, on the concentration-mass relation of galaxy systems. I built upon several observational results, showing that low-mass clusters and groups of galaxies are systematically over-concentrated with respect to theoretical predictions. Whenever possible, I adopted observationally motivated prescriptions as ingredients of the model, in order to partially bypass the uncertainty typical of numerical simulations stemming from the amount of cooling/feedback that is requested to reproduce a wide array of observable results. In a recent alternative approach, \citet{DU10.1} used a suite of cosmological simulations where different kinds of baryonic physics were implemented in order to study their impact on the structure of groups and clusters of galaxies. During the course of the paper I shall compare my results to those of \citet{DU10.1} where relevant.

The rest of this work is organized as follows. In Section \ref{sct:observed} I show the concentration-mass relation recently observed in different cluster catalogs by several authors. In order to allow a self-consistent comparison I accurately convert all masses and concentrations to the convention adopted in this work. In Section \ref{sct:model} I describe in detail the various ingredients entering the simple semi-analytic model that I employed in order to obtain a modified theoretical concentration-mass relation. In Section \ref{sct:results} I summarize the results, and in Section \ref{sct:discussion} I discuss them at length. Section \ref{sct:conclusions} is devoted to my conclusions. In what follows I adopted the standard flat $\Lambda$CDM cosmological model, with parameter values suggested by the latest WMAP data analysis \citep{KO11.1}. This implies a matter density parameter $\Omega_{\mathrm{m},0} = 0.272$, a cosmological constant density parameter $\Omega_{\Lambda,0} = 1-\Omega_{\mathrm{m},0}$, a Hubble constant $H_0 = h~100$ km s$^{-1}$ Mpc$^{-1}$, with $h = 0.704$, and a normalization of the matter power spectrum given by $\sigma_8 = 0.809$.

\section{Observed concentrations}\label{sct:observed}

In this Section I describe the concentration-mass relation resulting from several different observational studies of groups and clusters of galaxies. As mentioned in Section \ref{sct:introduction}, different authors in general adopt different conventions for defining the size (and hence the mass and concentration) of a structure, and in some occasion even slightly different cosmological parameter values. In order to make self-consistent comparisons it is hence necessary to convert all values of mass and concentration to the cosmology and convention adopted in this work. As a reminder, I define the size of a structure as the radius of the sphere whose average density equals to $\Delta\rho_\mathrm{c}(z)$, where $\Delta = 200$, and $\rho_\mathrm{c}(z) = 3H^2(z)/8\pi G$ is the critical density of the Universe at the redshift of the selected cluster. Please note that the critical density of the Universe and the average matter density are related via the matter density parameter $\Omega_\mathrm{m}(z)$, thus choosing the latter instead of the former as a reference density ultimately translates in a different value of $\Delta$ as a function of redshift. Analogously, different cosmologies produce a different $\rho_\mathrm{c}(z)$, which can also be regarded as an altered overdensity value as a function of redshift.

The procedure for converting masses and concentrations measured with respect to a certain overdensity $\Delta_1$ into those measured with respect to another overdensity $\Delta_2$ by assuming a NFW profile is relatively straightforward, and it is condensed in the Appendix of \citet{HU03.1}. Summarizing very briefly, the corresponding concentrations can be linked by numerically solving the algebraic equation

\begin{equation}
\frac{F(c_1)}{c_1^3} - \frac{\Delta_1}{\Delta_2}\frac{F(c_2)}{c_2^3} = 0~,
\end{equation}
(for which \citealt{HU03.1} give an approximate analytical solution) while the corresponding masses are linked by

\begin{equation}
M_1 = M_2 \left(\frac{c_1}{c_2}\right)^3\frac{\Delta_1}{\Delta_2}~.
\end{equation}
The function $F(x)$ reads

\begin{equation}
F(x) = \ln(1+x)-\frac{x}{1+x}~,
\end{equation}
and can be used to relate the mass with the concentration of a NFW dark matter halo according to $M = 4\pi\rho_\mathrm{s}r_\mathrm{s}^3F(c)$. It is easy to see that if $\Delta_1 > \Delta_2$, then $c_1 < c_2$, and also $M_1 < M_2$, because the radius of the structure must encompass a larger mean density.

In numerous of the observational works described below the authors refer to a modified concentration $c_0(m,z)\equiv (1+z)~c(m,z)$, rather than to the original concentration. The reason for this is the fact mentioned in Section \ref{sct:introduction} that according to some old theoretical concentration-mass prescriptions the concentration at a fixed mass is expected to scale with redshift as $\propto (1+z)^{-1}$. Thus, mutiplying the concentrations by $(1+z)$ would arguably remove this redshift dependence, and hence render measurements for objects at different redshifts more directly comparable. However, several more recent numerical studies have shown that the redshift dependence of the concentration is much weaker than that, and likely even vanishing for high-mass clusters. This has been verified observationally by \citet{SC07.2} (see below for more details) who found a redshift dependence of the concentration-mass normalization of $\propto (1+z)^{-0.71\pm 0.52}$. It is easy to understand that assuming a redshift dependence of the concentration significantly stronger than the actual one and then correcting for it would bias high the concentration of the highest redshift objects, that are presumably the most massive ones due to selection effects (especially in X-ray studies). This would have the effect of making the $c-M$ relation somewhat flatter than it actually is. Given this, and in agreement with \citet{OG11.1}, I chose not to include any kind of redshift correction, thus assuming that all the measured concentrations are as they would be at $z=0$. The weak redshift dependence of the concentration and the limited redshift range of the majority of the studies described below suggest that this is a good approximation, however I shall discuss more on it in Section \ref{sct:discussion}.

\subsection{The data}\label{sct:data}

I considered six of the most recent compilations of galaxy groups and clusters for the purpose of measuring the concentration-mass relation. In the following I describe each of them in detail, and summarize their findings. In Figure \ref{fig:concentration} I show the position (with relative errors) of objects in each catalog in the mass-concentration plane, as well as the expectation from the theoretical study of \citet{GA08.1} and the best fitting power-law for the observed clusters. Specifically, I fit the concentration-mass relation with the two-parameter function

\begin{equation}\label{eqn:pl}
\hat c(m,z=0~|~c_0,\alpha) = c_0 \left( \frac{m}{m_0} \right)^{-\alpha}~,
\end{equation}
where $m_0 = 5\times 10^{14}~h^{-1}M_\odot$, and the concentration for a fixed mass is assumed to be redshift-independent. In Figure \ref{fig:concentration}, the values of mass and concentration for each cluster have been carefully converted to the overdensity convention and fiducial cosmological model adopted in the present paper, as discussed above.

The observational data are fitted with the power law of Eq. (\ref{eqn:pl}), and the best fitting parameters are found by minimizing the $\chi^2(c_0,\alpha)$ function defined as 

\begin{equation}\label{eqn:chi2}
\chi^2(c_0,\alpha) =\sum_{i=1}^n \frac{\left[ \mathrm{Log}~c_i - \mathrm{Log}~\hat c(m_i,z=0~|~c_0,\alpha)\right]^2}{\sigma^2_i+\hat{\sigma}^2}~,
\end{equation}
where $n$ is the number of galaxy systems in the catalog, $\sigma_i$ is the logarithmic error on the measurement of the $i-$th concentration $c_i$, and $\hat\sigma$ is the expected intrinsic scatter in the concentration at a given mass, that we set to $\hat\sigma=0.15$ (see Section \ref{sct:introduction}). One should bear in mind that the best fitting normalization and slope depend on the chosen pivot mass, and that for an arbitrarily chosen $m_0$ the two shall have some degree of correlation. Nonetheless I am not interested here in finding the best possible fit to the data, rather only to demonstrate the degree of disagreement of the latter with theoretical models.

\subsubsection{\citet{BU07.1}}

In this work the authors used a sample of $39$ galactic systems spanning a wide range of masses, from isolated ellipticals ($m\sim 5\times 10^{12}~h^{-1}M_\odot$) to massive galaxy clusters ($m\gtrsim 10^{15}~h^{-1}M_\odot$). The objects were chosen to have the best quality \emph{Chandra} and \emph{XMM-Newton} observations to date. Moreover, in order to ensure hydrostatic equilibrium of the hot gas to be a good approximation, only systems displaying very regular X-ray maps, devoid of strong asymmetries, were selected. All objects in the sample are relatively local, having redshifts $z \lesssim 0.2$ and with only a handful having $z > 0.1$.

\citet{BU07.1} used the aforementioned X-ray observations to infer the spherically averaged mass profile of the objects in the sample, and hence determined the virial mass and the concentration for each one. It is worth stressing that in their analysis the authors corrected the concentrations by the extra factor $(1+z)$. As noted above, this possibly introduces a bias, making the concentration-mass relation flatter than it actually is. Despite this, they still found a slope of $\alpha = 0.17$, steeper than the theoretical expectation of $\alpha \sim 0.1$. Additionally, \citet{BU07.1} came across the fact that low-mass objects tend to lie systematically above the best-fitting relation, and that by including only massive systems in their statistical analysis the slope of the concentration-mass relation decreases substantially. These findings suggest that low-mass structure are responsible for the steepness of the relation.

The authors also discuss the normalization of the concentration-mass relation, explaining that it is consistent with the expectations for a $\Lambda$CDM cosmology having a high matter power spectrum normalization $\sigma_8=0.9$, although as explained previously low mass systems are systematically above that. This, and the fact that recent observations favor a $\sigma_8$ value actually closer to $0.8$ rather than $0.9$ \citep{KO11.1}, would advocate again for some mechanism to increase halo concentrations with respect to the predictions of pure $n-$body simulations. The top left panel of Figure \ref{fig:concentration} summarizes the situation for this catalog. Without correcting for any redshift dependence, I obtain a best fit slope of $\alpha = 0.20\pm0.04$, marginally steeper than the one obtained originally by \citet{BU07.1}. The theoretical prediction by \citet{GA08.1} is significantly flatter than required by the data, with only $4$ out of $39$ observed structures lying below it.

\subsubsection{\citet{CO07.1}}

In this work the authors presented a compilation of virial mass and concentration measurements derived in different works and using different techniques. Specifically, the mass distributions of $100$ galaxy clusters and groups were obtained using X-ray observations, gravitational lensing (weak and/or strong), kinematic analysis, i.e., line-of-sight velocity dispersions, and the caustic method. The fact that different techniques are mixed together is a significant variant with respect to other works, where commonly all cluster parameters are measured adopting the same technique. This makes it more difficult to interpret the results, as different methods are (or may be) affected by different biases. The cluster sample spans a mass range included between $m \sim 3\times 10^{13}~h^{-1} M_\odot$ and $m \sim 2\times 10^{15}~h^{-1} M_\odot$, and comprises very local systems as well as a few distant ones (up to $z\sim 0.9$).

Similarly to \citet{BU07.1}, the authors multiplied the cluster concentrations by the extra factor $(1+z)$, thus possibly biasing the slope of the concentration-mass relation low. According to their analysis, this slope measures $\alpha = 0.15$, consistent with the \citet{BU07.1} result and marginally steeper than the theoretical expectation. Also, the normalization of the relation is substantially higher than, and basically in disagreement with, the one predicted for a cosmology with $\sigma_8 \sim 0.8$. In the top right panel of Figure \ref{fig:concentration} I report the concentration-mass relation for the catalog collected by \citet{CO07.1}. Similarly to what they did, I also removed all objects that have no mass measurements or no reported errors for either the mass or the concentration. When more than one measurement is available for the same cluster, I use the average values, with errors equal to the largest errors reported. The observed scatter is very large, resulting in a slope of $\alpha = 0.12\pm0.06$, consistent with both the theoretical prediction of \citet{GA08.1} and the original analysis of \citet{CO07.1}. The normalization on the other hand is substantially larger than expected for a WMAP-based cosmological model.

\subsubsection{\citet{ET10.1}}

These authors used X-ray surface brightness and temperature profiles measured by \emph{XMM-Newton} for a sample of $44$ luminous galaxy clusters at $0.1\lesssim z \lesssim 0.3$ to infer their mass distributions, and hence the relation between the virial mass and the concentration. \citet{ET10.1} excluded from their sample objects showing signs of recent and strong interactions, in order to ensure the hydrostatic equilibrium to be a good approximation for the hot gas.

They used two different methods in order to reconstruct the mass profiles of clusters in their samples, which give consistent results between each other. I shall refer to the results of their first method, as it has been extensively tested against numerical simulations. In this work as well the authors corrected each concentration value by the factor $(1+z)$, and found that both the slope and the normalization of the relation agree with expectations from numerical simulations. However, they also claim that the range in mass they explore, $m \gtrsim 10^{14}~h^{-1} M_\odot$ is too small to draw any definitive conclusion. 

\begin{figure*}
	\includegraphics[width=0.43\hsize]{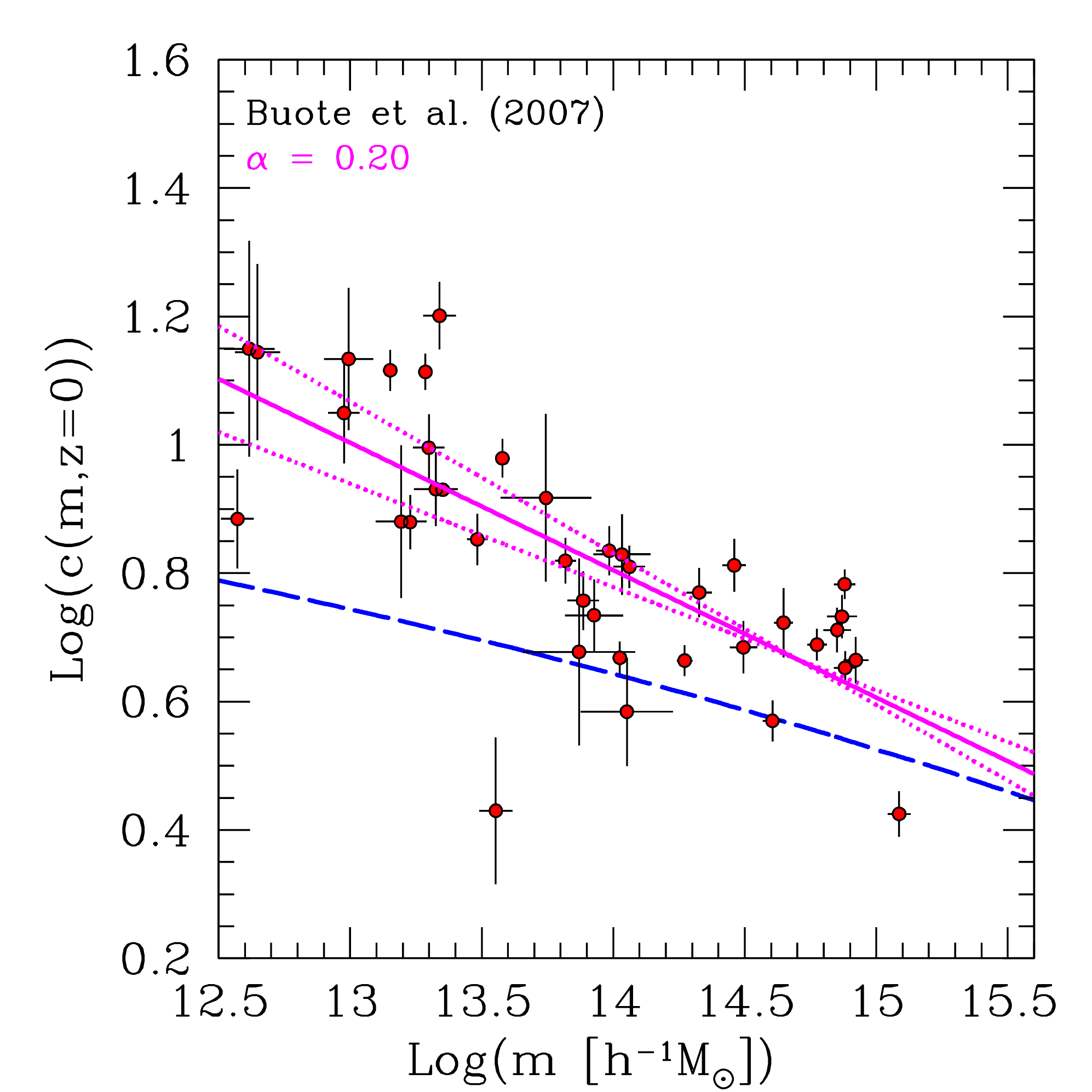}
	\includegraphics[width=0.43\hsize]{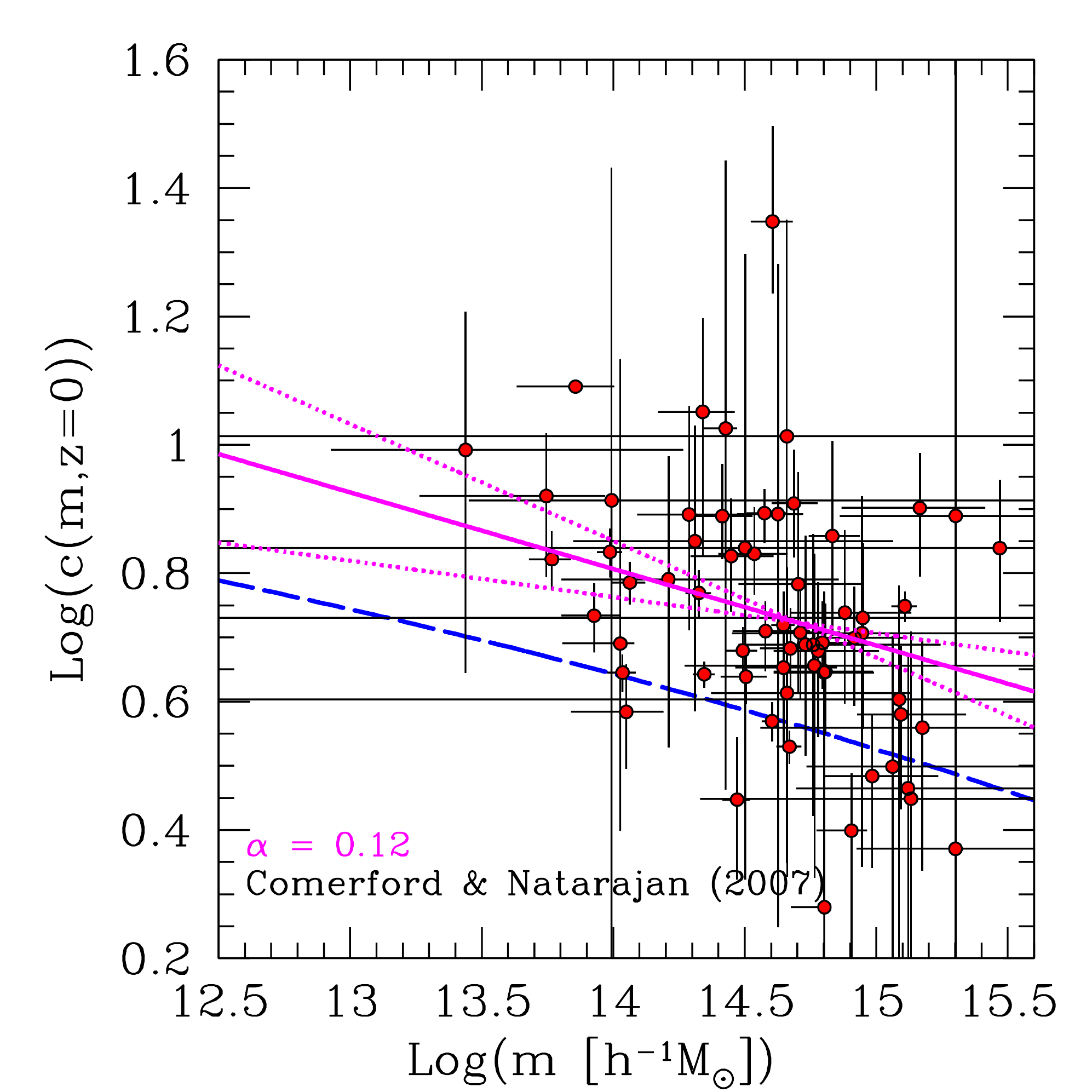}
	\includegraphics[width=0.43\hsize]{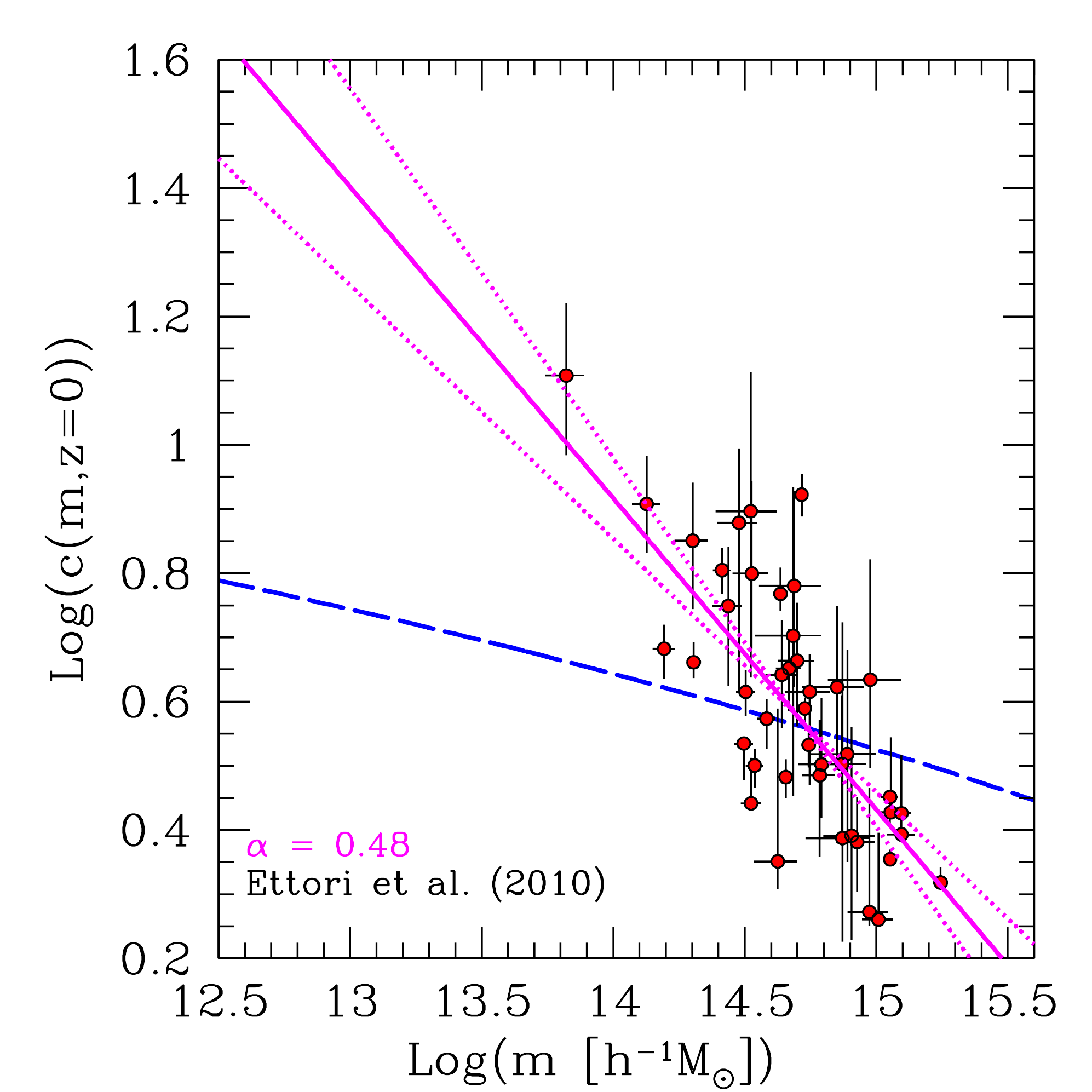}
	\includegraphics[width=0.43\hsize]{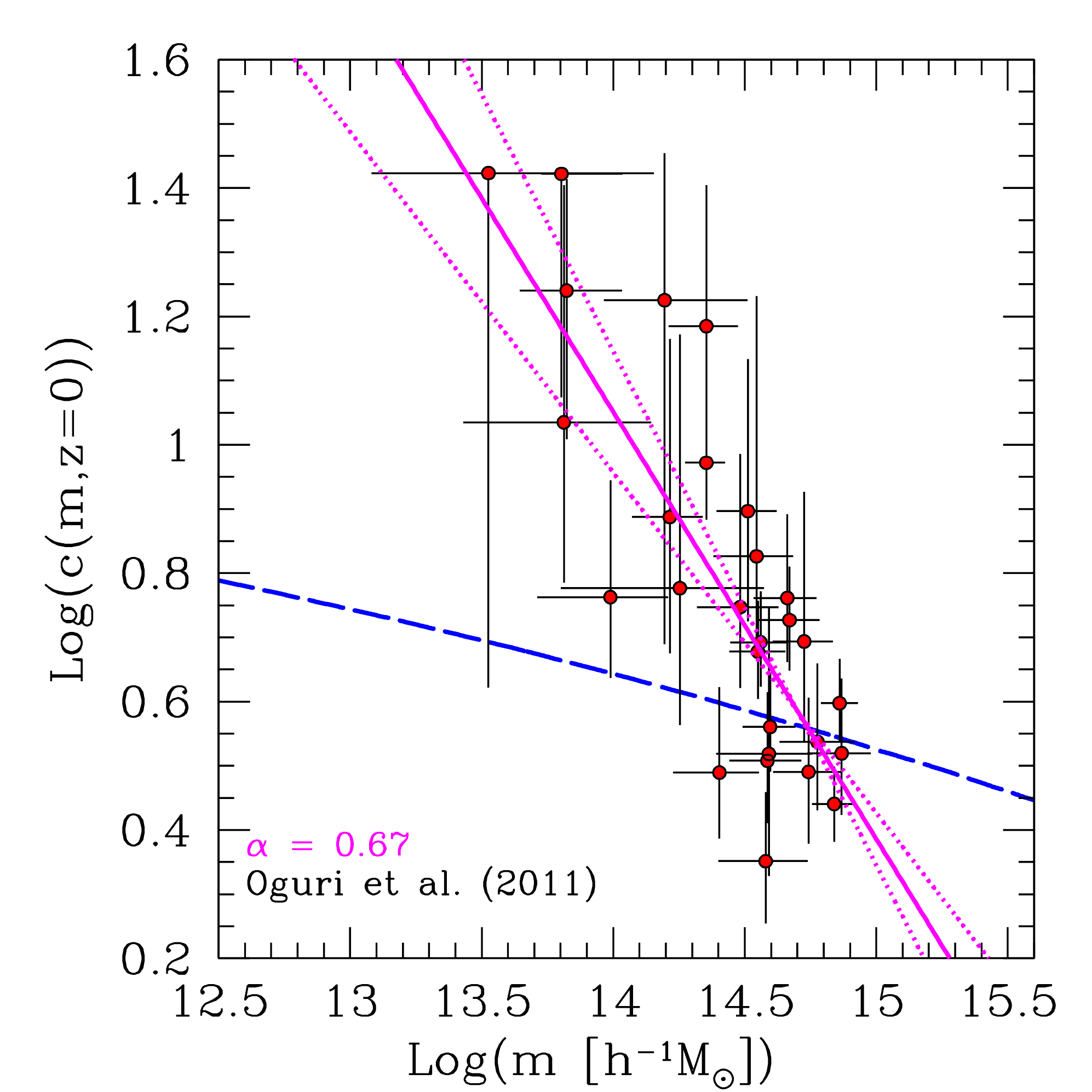}
	\includegraphics[width=0.43\hsize]{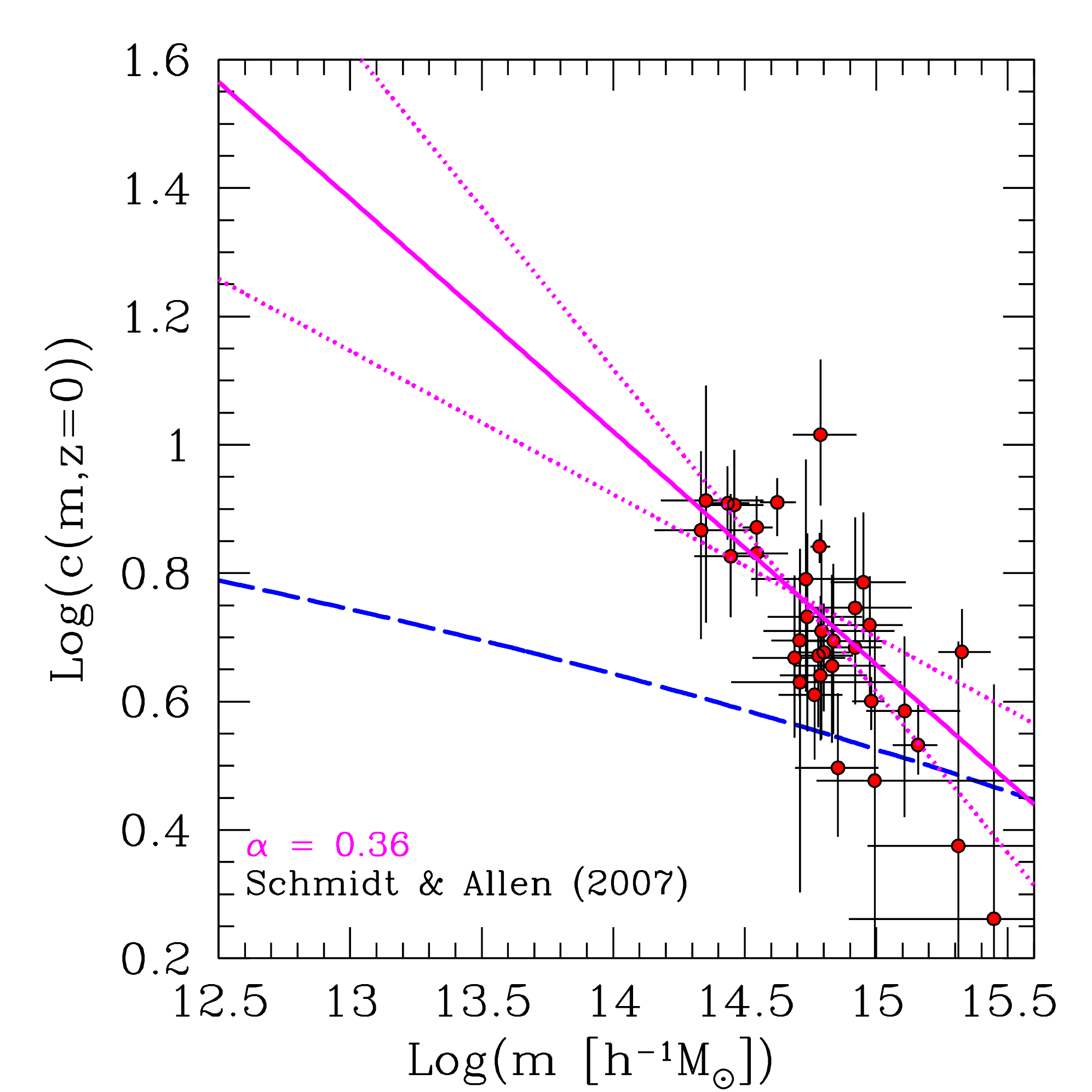}
	\includegraphics[width=0.43\hsize]{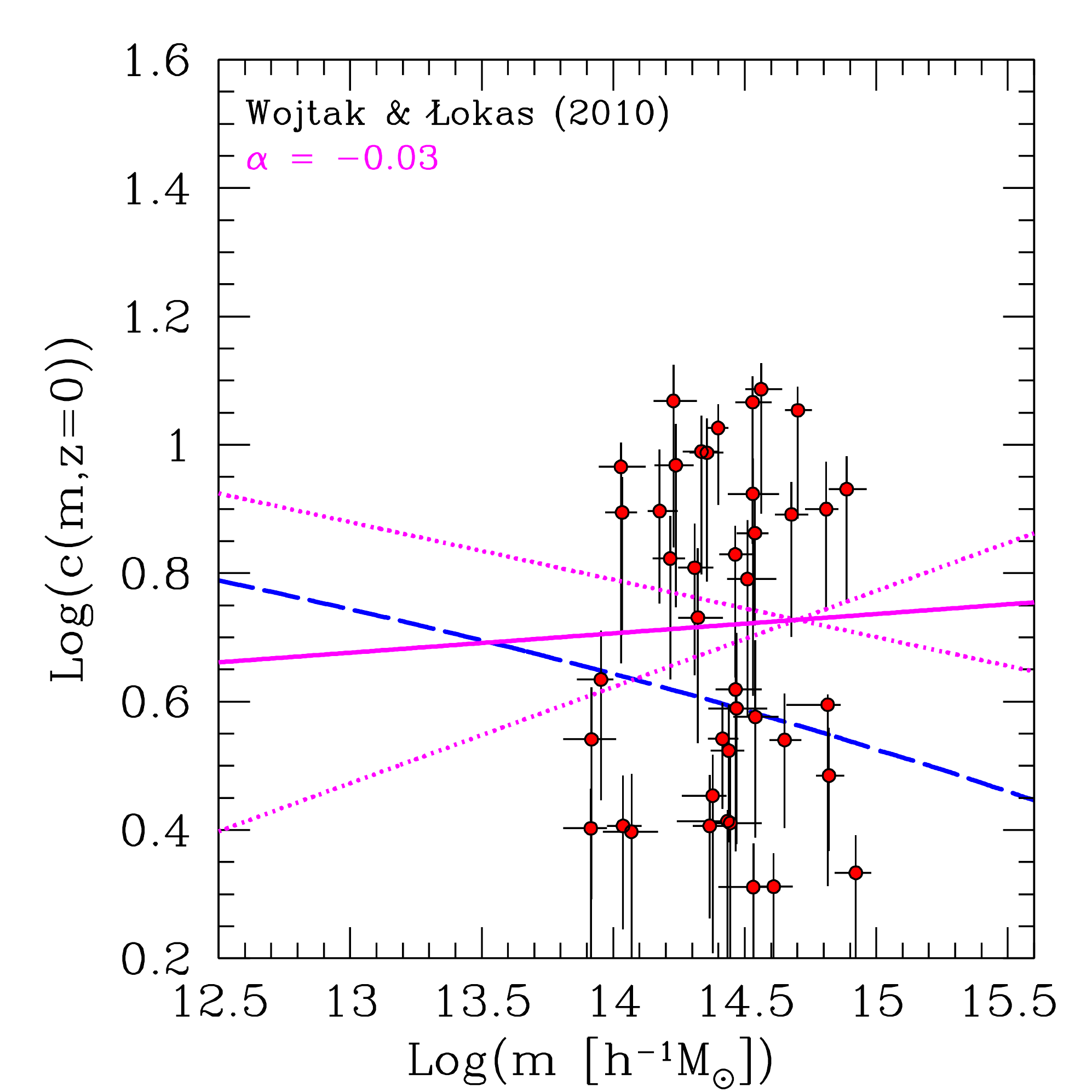}
	\caption{The observed relation between the concentration and the mass for groups and clusters of galaxies. Each panel refers to a different catalog of objects, compiled by the labeled authors. The blue long-dashed line represents the concentration-mass relation predicted by Gao et al (2008) at $z = 0$, while the solid magenta line is the best fit to the observed points, with labeled slope. The dotted magenta lines show the uncertainty on the best fitting slope.}
\label{fig:concentration}
\end{figure*}

In the middle left panel of Figure \ref{fig:concentration} I report the concentration-mass relation as measured for the sample of \citet{ET10.1}. The best fitting slope is $\alpha = 0.48\pm0.09$, substantially larger than the theoretical expectation and of the original finding of \citet{ET10.1}. This difference is likely due to the $(1+z)$ factor correction that they perform and to the different fitting techniques adopted. It should be noted that, contrary to the other works I consider here, the mass and concentrations of \citet{ET10.1} refer to the dark matter only distribution rather than to the total mass. It is something to keep in mind for the interpretation of the results, however it does not change my conclusions, as discussed in Section \ref{sct:discussion}.

\subsubsection{\citet{OG11.1}}

Very recently, these authors used combined weak and strong lensing information in order to estimate the mass profiles of a sample of $28$ galaxy clusters selected  for having bright gravitational arcs in the SDSS cluster sample. The resulting catalog covers a rather large range in virial mass, $3\times 10^{13}~h^{-1} M_\odot \lesssim m \lesssim 10^{15}~h^{-1} M_\odot$, and the redshift range where gravitational lensing is typically more efficient for sources at $z_\mathrm{s}\sim 2$, that is $0.3 \lesssim z \lesssim 0.6$.

\citet{OG11.1} acknowledge the recently established lack of redshift evolution for the concentration at a given mass (at least for the largest objects, \citealt{GA08.1,PR11.1,ZH09.2}), and therefore do not correct the measured concentrations according to any redshift dependence. They find a surprisingly high slope for the concentration-mass relation, $\alpha \sim 0.6$, substantially higher than that inferred from numerical simulations. Very interestingly, the authors propose a modification of the theoretical concentration-mass relation that takes into account the strong lensing bias, namely the fact that strong lensing systems have on average a larger intrinsic concentration, and are preferentially elongated along the line of sight with respect to the general cluster population. Since this bias is more marked for low-mass systems, this increases the slope of the concentration-mass relation up to $\alpha \sim 0.2$, still much lower than found observationally. 

\citet{OG11.1} identify the problem as lying in the low-mass systems, as the high mass ones seem to be quite compatible with their model. This is a similar conclusion to \citet{BU07.1}. In the middle right panel of Figure \ref{fig:concentration} I show the \citet{OG11.1} sample, compared with the theoretical expectations. The observed slope that I find is $\alpha = 0.67\pm0.13$, in good consistency with the original estimate of the authors, and the highest slope of all the samples considered here. This latter fact can be ascribed to this being the only purely strong lensing selected sample, in which the lensing bias must play a role (although not sufficient to reconcile the observed slope with theoretical predictions).

\subsubsection{\citet{SC07.2}}

In this work the authors used \emph{Chandra} X-ray observations of $34$ massive and relaxed galaxy clusters in order to infer the respective mass distributions. The mass range they explored is relatively narrow, being limited to rich groups and massive clusters, $m \gtrsim 3\times 10^{14}~h^{-1} M_\odot$. The studied systems display a variety of redshifts, ranging from almost local objects up to very distant ones ($z\sim 0.7$). \citet{SC07.2} managed to separate the dark matter halo mass profile from the distribution of the luminous matter components, however in order to compare with other studies, they also presented results of fitting the total matter distribution with a single NFW profile, which I adopted here.

They compared their observational results with a power-law in mass having parameters estimated from numerical cosmological simulations, and in particular having a normalization that depends on redshift as $\propto (1+z)^{-1}$. Their results show that this model substantially underestimates the concentration of low-mass systems, thus providing an overall poor fit to the observed concentration-mass relation. On the other hand, by relaxing the redshift dependence of the concentration normalization and allowing the slope of the concentration-mass relation to be free as well, they found a significantly better fit, having both a slope substantially higher than predicted by theoretical models ($\alpha = 0.45$), and a remarkably small redshift dependence $\propto(1+z)^{0.71\pm 0.52}$ (interestingly in agreement with the simulations of \citealt{DU08.1}). All in all, this confirms the steep slope of the concentration-mass relation found in other works, and the mild redshift dependence of the concentration at fixed mass derived from the most recent numerical studies. In the bottom left panel of Figure \ref{fig:concentration} I show the results of \citet{SC07.2}, together with their best fitting power law. I obtain a slope of $\alpha = 0.36\pm0.14$, consistent with the original estimate of the authors. This value is much larger than the expectations of theoretical investigations.

\subsubsection{\citet{WO10.1}}

In this last work, the authors employed kinematic data for a sample of $41$ nearby ($z\lesssim 0.1$) and relaxed galaxy clusters with $m\gtrsim 10^{14}~h^{-1}M_\odot$, in order to constrain their mass profiles. The relaxation of a system was evaluated on the basis of X-ray observations and the line-of-sight velocity structure, specifically, all clusters having strongly asymmetric or substructured X-ray maps, as well as bimodal velocity distributions were discarded from the sample.

Contrary to the previous works summarized in this Section, \citet{WO10.1} found a slope of the concentration-mass relation consistent with flatness. On the other hand, the measured normalization is somewhat higher than predicted by theoretical models, resulting in a marginal inconsistency with a WMAP-based standard cosmology. In the bottom right panel of Figure \ref{fig:concentration} I report the results of \citet{WO10.1}. I also found consistency with a flat concentration-mass relation ($\alpha=0.03\pm0.12$), however the observed data do not seem to be in strong disagreement with the \citet{GA08.1} prescription, although the measured concentrations seem to be somewhat systematically larger than theory would suggest.

\subsection{Implications}

The scrutiny of previous works presented in Subsection \ref{sct:data} above shows that in general the concentration-mass relation of observed clusters and groups of galaxies is substantially steeper than any theoretical prediction. In the two cases in which the measured slope is compatible with the prediction of the \citet{GA08.1} prescription, namely for the samples of \citet{CO07.1} and \citet{WO10.1},  the normalization results too high to be in agreement with a $\Lambda$CDM cosmological model having $\sigma_8\sim 0.8$, as suggested by the latest WMAP data. Interestingly, this conclusion remains true quite irrespectively of the cluster selection function and analysis procedure, holding for either strong lensing clusters and for X-ray selected ones. In the former case one actually expects some level of discrepancy with respect to theory, since strong lensing clusters are usually biased to have a larger inferred concentration than the average of the entire population, and the bias tends to be stronger for lower mass systems (see \citealt{HE07.2,ME10.1}). In their recent work, \citet{OG11.1} proposed a simple scheme in order to account for this lensing bias, finding a somewhat larger normalization of the concentration-mass relation and a slope that however can be increased up to $\alpha \simeq 0.20$ at the most. While this would probably be a better fit to the data of \citet{CO07.1}, it is way insufficient to match their own data. Plus, X-ray selected clusters and kinematic studies are not subject to any lensing bias, and still often show similar discrepancies.

It should be noted that the observational works considered in this paper are by no means the only ones on the subject. There are plenty of them that are either older or making use of more limited cluster samples. Despite their conclusions being less strong, many of these works hint at discrepancies similar to those highlighted above. An example is the work of \citet{GA07.1}, whose data have been recycled in \citet{BU07.1}. I also refer the reader to \citet*{MA08.2} for a paper finding agreement with the theoretical concentration-mass relation.

The systematicity with which observed low-mass systems are over-concentrated with respect to the theoretical expectations calls for some mechanism that is ignored in $n$-body simulations and that would be more effective at group scale rather than cluster scale. In this work I investigated the most obvious ingredient, that is the physics of baryons. The presence of gas and stars in real structures affects estimates of concentrations in a twofold way: $i)$ observations often infer the total mass distribution of galaxy systems, which includes dark matter, hot diffuse gas, and stars. Numerical simulations on which theoretical models are based instead follow only the evolution of dark matter; $ii)$ the very cooling of gas in the central regions of dark matter halos imply a contraction of the dark matter component itself, an effect that is dubbed baryonic contraction (see \citealt{GN11.1} and references therein). Given that gas cooling and star formation are expected to be more efficient in low-mass objects, both these effects should be more enhanced in those systems. It is my aim to make use of a simplified spherical model that takes into account these two factors, in order to see if the impact of baryons can indeed account for the observed steepness/normalization of the concentration-mass relation.

It is worth to be noted that in several of the catalogs considered here there are a few objects having significantly lower concentrations than the theoretical expectations. The same has been found in the weak lensing analysis of \citet*{MA08.2}. It is quite obvious that the gas cooling and star formation have only the effect of increasing the concentration of a structure, hence they cannot account for those data. Strong energy feedback from AGN as well as observational biases might be at work for those structures, which I shall discuss better in Section \ref{sct:discussion}.

\section{A simple spherical model}\label{sct:model}

Baryons are present in two phases inside their host dark matter halos. The first phase is a hot diffuse gas component, which is responsible for the emission of thermal X-ray radiation. This component dominates the baryon budget in rich groups and clusters of galaxies. The second phase is made by the cold gas and stars that constitute galaxies, and that result from the cooling and condensation of the hot phase in compact regions. This phase is dominant in small groups and individual galaxies, since their shorter cooling time makes it easier to form massive stellar clumps.

\subsection{The baryon fraction}\label{sct:fraction}

The first fundamental ingredient for any self-consistent cosmic structure model lies in the fractions of hot gas and cold gas/stars that are present in their host dark matter halo, as a function of its mass and redshift. This issue is the focus of much active investigation on the observational side, and a clear consensus has not been achieved yet. As of today it seems quite clear that the majority of baryons is outside bound structures, in the form of a diffuse and smoothly distributed gas component \citep{BR07.1}. However the actual fraction of baryons within dark matter halos and especially their partitioning between hot gas and stars is the subject of substantial debate. In what follows, I shall introduce the prescription that has been adopted in this work, and then discuss at length how this compares with other works in the same field. 

\begin{figure}
	\includegraphics[width=\hsize]{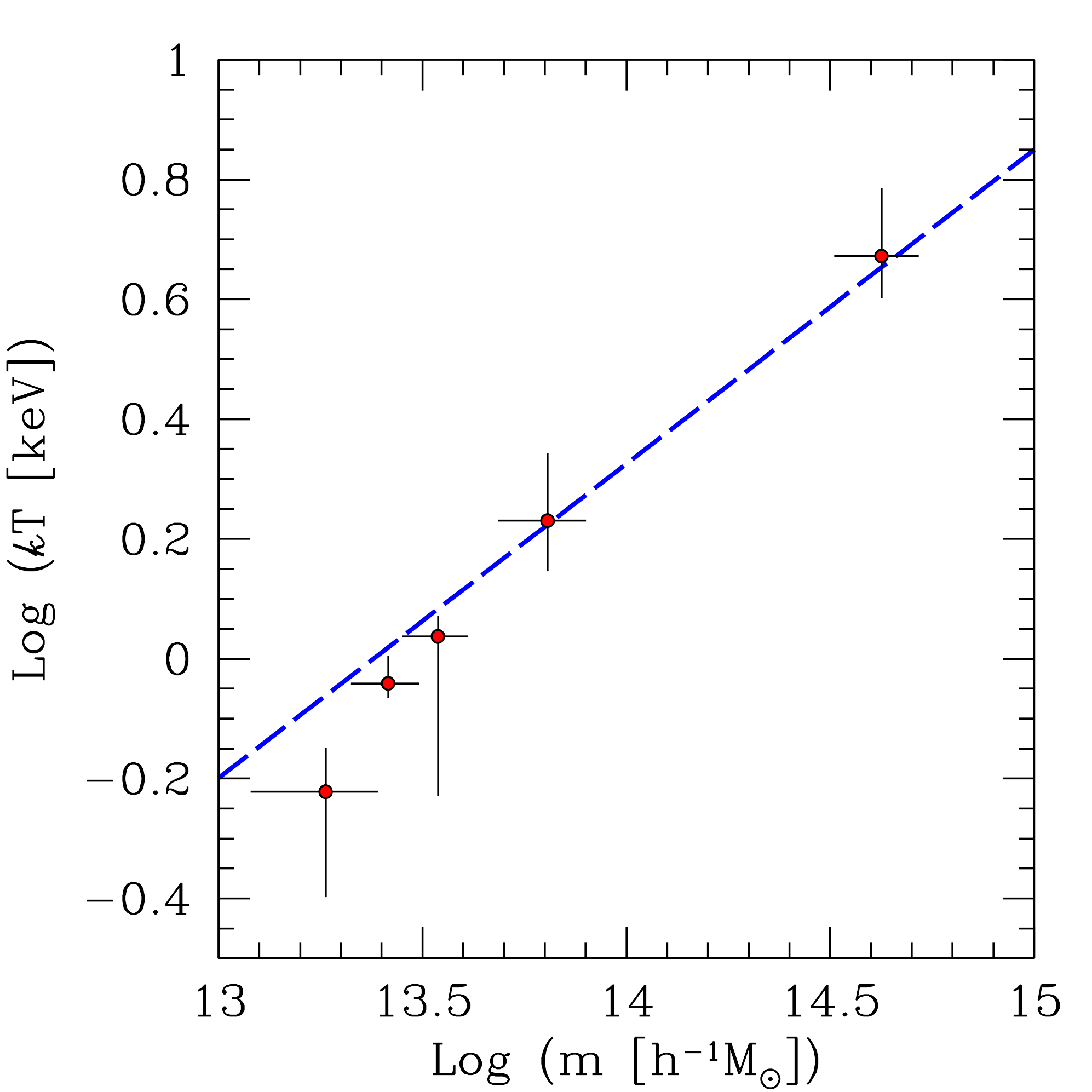}
	\caption{The temperature-mass relation. Red circles with errorbars represent the results of the stacking analysis performed by \citet{DA10.1} in five different richness bins, ranging from poor galaxy groups to relatively rich galaxy clusters. The blue dashed line is the scaling law obtained by combining the observed $\sigma-T$ relation of \citet{XU00.1} with the $m-\sigma$ relation of \citet{EV08.1} at $z=0$.}
\label{fig:massTemperature}
\end{figure}

In the recent paper by \citet{DA10.1} the authors employ stacking techniques in order to derive the fraction of both hot gas and stars out to the virial radii of the host galaxy groups. With respect to other investigations on the same topic, this work has a twofold advantage from my point of view. First, it combines the stacked data for galaxy groups to pre-existing data on clusters as well as massive and dwarf galaxies, thus covering a very wide dynamical range. The majority of other works focus instead on very limited mass ranges. Given that the concentration data I am considering cover $\sim 3$ orders of magnitude in mass, this is a desirable feature. Second, \citet{DA10.1} refer where possible their baryon fractions to the virial radii of structures (defined in the same way as done in the present work), while virtually all other works refer to radii encompassing overdensities of $500$ or more with respect to the critical density of the Universe. This is another positive aspect, since many of the theoretical and observational works on concentrations refer to virial quantities instead. Consequently, I adopted the fit to the baryon fractions measured in the work by \citet{DA10.1}, referring to that paper for additional details.

There is one caveat though, namely \citet{DA10.1} refer their results to the X-ray temperature or velocity dispersion of the structure at hand, thus forcing one to adopt scaling relations in order to convert them into a mass. As the authors themselves do, I convert the temperature into a velocity dispersion by using the experimentally calibrated relation of \citet{XU00.1}, namely

\begin{equation}
\sigma = 309~\mathrm{km~s}^{-1}~\left( \frac{kT}{1~\mathrm{keV}} \right)^{0.64}.
\end{equation}
Then, by assuming that galaxies and dark matter particles are in dynamical equilibrium, I convert the velocity dispersion into a mass by adopting the simulation-based scaling law derived by \citet{EV08.1}. The combination of these two relations agrees reasonably well with the result of the stacked analysis in the five richness bins adopted by \citet{DA10.1} themselves, that cover the group-cluster mass scale. This is reported in Figure \ref{fig:massTemperature}. As can be seen, the resulting observed temperature-mass relation tends to become somewhat steeper toward low-mass structures than it is at cluster scale. I tried to steepen a bit the association for low-mass groups, finding that the result is a substantial drop in the baryon fraction for $m \lesssim 10^{13}~h^{-1} M_\odot$, a regime that is basically of no interest here.

\begin{figure}
	\includegraphics[width=\hsize]{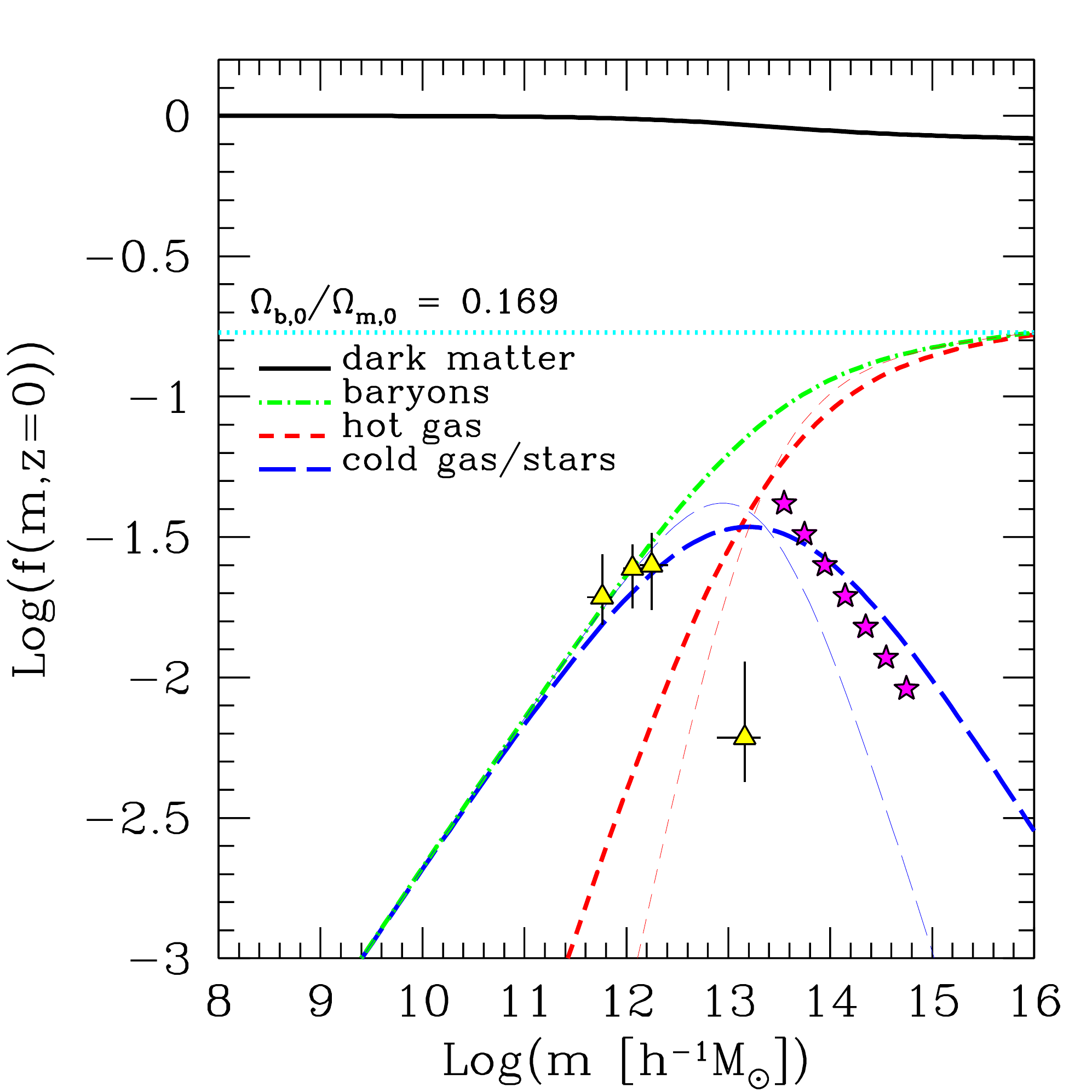}
	\caption{The baryon fraction as a function of total mass for $z=0$, as given by the fit of \citet{DA10.1}. Each line style and color refer to a different matter component, as labeled in the plot. The horizontal cyan dotted line represents the universal baryon fraction given by the WMAP$-7$ cosmological parameters adopted in this work. The yellow triangles represent the stellar fractions measured at $z=0$ by \citet{CO07.2}, while the magenta stars show the same quantity as measured by \citet{AN10.1}. Errorbars are not shown here since they are of the same size of the points. The thin lines show the gas and stellar mass fractions obtained by changing one of the parameters of the model, as explained in detail in the text.}
\label{fig:massFractions}
\end{figure}

The baryon fraction resulting from the analysis of \citet{DA10.1}, combined with the scaling relations mentioned above is depicted in Figure \ref{fig:massFractions}. As can be seen, the stellar fraction increases monotonically with mass up to $m\sim 10^{13}~h^{-1}M_\odot$, and then it starts a rapid decrease. That is also the same mass scale at which the gas fraction becomes as important as the stellar fraction, surpassing the latter for larger masses and being subdominant for smaller ones. The total baryon fraction always grows with mass, however it flattens at the mass scale of groups and clusters of galaxies, reaching the universal baryon fraction only for super-clusters. To my knowledge there are only two other works that push the measurement of the stellar mass fraction (or baryon fraction in general) out to the virial radius, that are \citet{CO07.2} and \citet{AN10.1}. For this reason, these findings are the only ones that can possibly be compared directly to the fit of \citet{DA10.1}. 

The results of \citet{CO07.2} are represented in Figure \ref{fig:massFractions} by the yellow triangles with errorbars, while the results of \citet{AN10.1} are shown as magenta stars. As can be seen, the latter data are in rather good agreement with the prescription adopted here, although the stellar mass fraction seems to evolve somewhat more steeply with mass. The data of \citet{CO07.2} also are in agreement with the prescription of \citet{DA10.1}, except for the highest mass point, which is lower by a factor of $\sim 4$. I note that the data points by \citet{CO07.2} do not take into account the stellar mass contribution of satellite galaxies, hence it is likely that the total stellar mass fraction displayed by the rightmost yellow triangle be significantly underestimated. It should also be kept in mind that the different works mentioned above (and below) adopt different conversions between luminosity and stellar mass, which can easily lead to $\sim 50-100\%$ discrepancies in the stellar fraction. Since I am only interested to gauge the possible impact of baryons on the concentration-mass relation, performing a highly detailed analysis of these conversions is beyond the scope of the paper. I will hence adopt the \citet{DA10.1} prescription as the fiducial one here, recalling that significant fluctuations around it are allowed.

Several additional works studied the issue of the baryonic content within cosmic structures, both from the points of view of the stars and of the hot diffuse gas. As mentioned above, all these works refer to substantially smaller radii than the virial radius, hence in order to make a fair comparison with the model of \citet{DA10.1} one should convert the masses (assuming specific profiles) and extrapolate somehow the baryon fractions. This latter operation is highly uncertain, as the distribution of baryons in the outskirts of clusters (clumpiness, depletion, etc., see for instance \citealt{EC11.1}) is still very poorly understood. Therefore I do not attempt here such a procedure, rather I discuss only the most generic features of the baryon fraction, such as the slope as a function of mass, the scale of turn-around for the stellar mass fraction, etc.

As already alluded to, the results on the stellar mass fraction of \citet{AN10.1} are in fair agreement with the model adopted here. The author also combined these findings with gas mass fractions measurements by \citet{VI06.1} and \citet{SU09.2}, and upon correction for the different radial coverage adopted they concluded that the stellar and gas mass fractions are equal at $m \sim 10^{13}~h^{-1} M_\odot$, in agreement with Figure \ref{fig:massFractions}. Also, the total baryon fraction is a very shallow function of mass at cluster scales, again in agreement with the model adopted here. In a less recent X-ray-based analysis, \citet{GA07.1} found a significant decrement of the fraction of diffuse gas in galaxy groups as compared to cluster-scale objects, as can be seen in the fit of \citet{DA10.1} depicted in Figure \ref{fig:massFractions}.

In \citet{GI09.2} the authors studied the baryon fraction of a large sample of X-ray selected groups. They concluded that, when extending their mass range by considering previously analyzed galaxy clusters, the stellar fraction scales with the mass with a power-law exponent of $-0.37$, compatible with the $\sim -0.4$ that I find at the high-mass end for the model adopted here. In the same work, the gas fraction is also studied, based on previously existing compilations. The result is that the gas fraction increases with increasing mass, although the slope is quite modest ($\sim 0.2$). Finally, \citet{GI09.2} also found a significant increment of the total baryon fraction as a function of mass at cluster scales, in agreement with Figure \ref{fig:massFractions} of this paper. However, their slope is only of $0.09$, and they admittedly do not include the contribution from the Intra-Cluster Light (ICL), which might flatten it further. The latter contribution has been studied by \citet*{GO07.2}. These authors found that the stellar mass fraction changes somewhat more markedly with mass than found by \citet{GI09.2} and \citet{DA10.1}, while the change in diffuse gas fraction at cluster mass scales is compatible. Interestingly, the slope for the change of the gas mass fraction with mass ($\sim 0.21$) is also consistent with the one found by \citet{PR09.1}. Most remarkably, \citet*{GO07.2} found that the total baryon fraction is constant for massive groups/clusters, advocating the impact of the ICL for this discrepancy with other works.

In a different study, \citet{BA07.1} investigated the stellar mass fraction for a sample of galaxy groups, extending their analysis out to the virial radius. They found a definite decrement in the latter as a function of halo mass, however the uncertainties displayed in their results are very large (ranging from a factor of $5$ up to one order of magnitude), and for this reason I did not consider them in Figure \ref{fig:massFractions}. Nevertheless, those results seem to point toward a steeper decrease of the stellar fraction as compared to the model adopted here, consistently with \citet*{GO07.2}.

I conclude this Subsection by considering two very recent works that return results very different from the bulk of previous investigations with respect to the stellar mass fraction. Specifically, \citet{LE11.1} found a content of stars for a sample of X-ray galaxy groups substantially lower than previous estimates (factors between $2$ and $5$). Their results are marginally consistent with those of \citet*{BE10.1}, that return even lower values for the stellar mass fraction. Interestingly the highest mass point of \citet{CO07.2} (which misses the contribution of satellites) is compatible with these works. Another difference between these results and those derived by, e.g., the \citet{DA10.1} analysis lies in the fact that the turn-around of the stellar mass fraction is at substantially lower masses, $\sim 5\times 10^{11}~h^{-1}M_\odot$. It is not yet clear what is the main reason behind these largely discrepant results. \citet{LE11.1} advocate inaccurate estimators for the stellar masses in previous works, however some problem might as well reside in the abundance matching technique adopted by these authors.

\begin{figure*}
	\includegraphics[width=0.43\hsize]{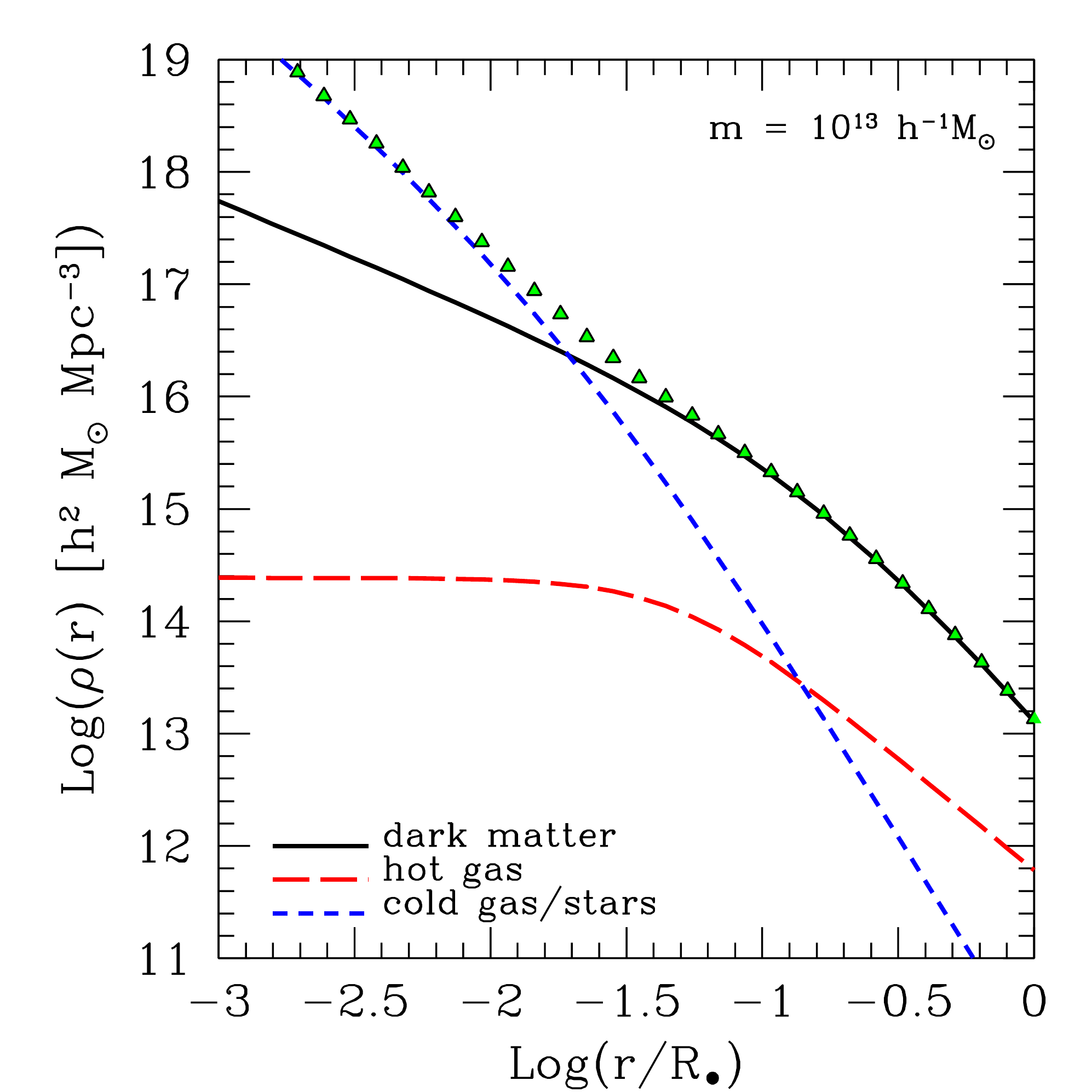}
	\includegraphics[width=0.43\hsize]{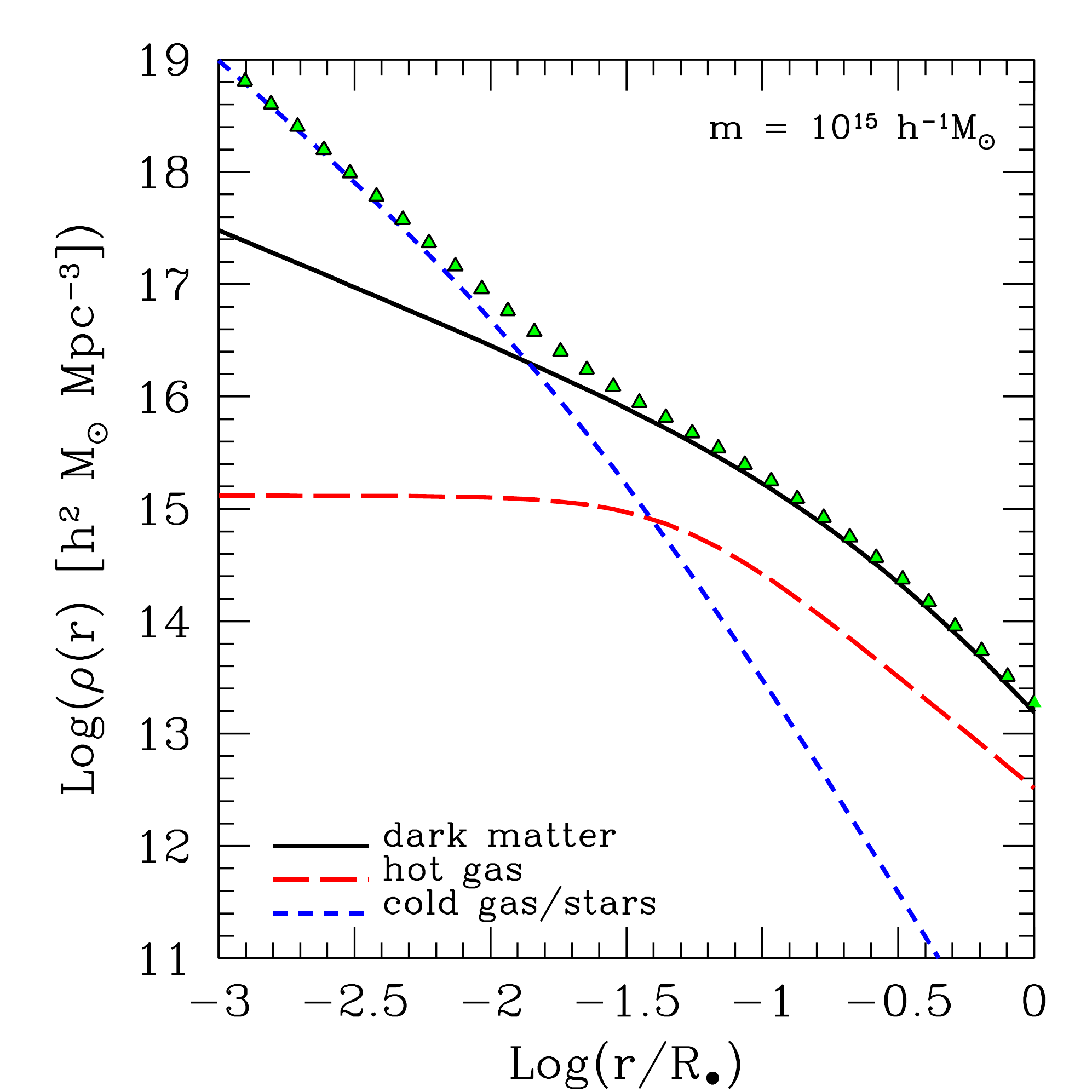}
	\caption{The density profiles of different matter components in a system with total mass $m = 10^{13}~h^{-1}M_\odot$ (left panel) and one with total mass $m = 10^{15}~h^{-1}M_\odot$ (right panel). Each line refers to a different component, as labeled, while the green triangles show the total mass density profiles.}
\label{fig:densityProfiles}
\end{figure*}

In summary, excluding the works by \citet{LE11.1} and \citet*{BE10.1}, the baryon fraction derived by the data of \citet{DA10.1} in combination with suitable scaling relations is in good qualitative agreement with other works on the topic. Possible indications of a steeper slope for the stellar content (as found by \citealt{BA07.1}, \citealt*{GO07.2}, and \citealt{AN10.1}) are only marginal and likely irrelevant for our conclusions. A possible discrepancy is the finding of \citet*{GO07.2} that the total baryon fraction should be constant with mass at cluster scales. Although the slight increase found by, e.g., \citet{AN10.1} is consistent with those indications, this has not been stressed by other authors, and it might also be due to the different radii at which mass fractions are referred. In order to accommodate for a substantial reduction in the amount of stars in high-mass structures, as advocated recently by \citet{LE11.1}, I multiplied by $2$ the slope of the correlation between the stellar to gas mass ratio and the virial temperature presented in \citet{DA10.1}. The resulting gas and star mass fractions are shown as thin lines in Figure \ref{fig:massFractions}. While keeping the original fit of \citet{DA10.1} as the fiducial model, I will later show results for this latter configuration as well.

\subsection{Matter distribution within spherical systems}\label{sct:distribution}
 
Given the mass fractions of the different matter components that are present inside any given object, it remains to be established how these components are effectively distributed, i.e., their density profiles. Before looking into that, let me set up a little of notation. I shall indicate with $f_\mathrm{g} (m,z)$ the fraction of hot diffuse gas present in a structure with total mass $m$ placed at redshift $z$. Likewise, $f_\star(m,z)$ and $f_\bullet(m,z)$ represent the fractions in stars and dark matter for the same structure. It is obvious that $f_\bullet(m,z) + f_\mathrm{g}(m,z) + f_\star(m,z) = 1$. 

Now back to the density profiles, the dark matter distribution prior to the contraction due to baryonic cooling shall be represented by a NFW density profile, with the concentration-mass relation suggested by \citet{GA08.1}. The only caution to be exerted is that the mass of the dark matter halo (to which the concentration is attached) is not the total mass $m$ of the structure at hand, rather it is $f_\bullet(m,z)~m$. Also, it should be kept in mind that the cooling of gas and the formation of stars cause the dark matter mass distribution to become more compact than predicted by pure $n$-body simulations, so that the final dark matter density profiles will not be NFWs anymore. I shall go back to this issue later below in Subsection \ref{sct:contraction}.
 
For the hot diffuse gas component I adopted a $\beta$-profile \citep{CA76.1}, which reads
 
\begin{equation}
\rho_\mathrm{g}(y) = \frac{\rho_\mathrm{c}}{(1+y^2)^{3\beta/2}}~,
\end{equation}
where $y\equiv r/r_\mathrm{c}$. This cored profile has formally three parameters, the outer slope $\beta$, the core radius $r_\mathrm{c}$, and the corresponding core density $\rho_\mathrm{c}$. The latter however acts just as a normalization, whose value is determined by matching the total gas mass within the structure with the gas fraction defined in the previous Subsection \ref{sct:fraction}. I shall assume as a characteristic object size the virial radius of the dark matter component prior to baryonic contraction, labeled as $R_\bullet$. This is very similar to the virial radius of the structure as a whole, having the advantage of being independent on the distribution of baryons. Hence, the following relation for the hot gas is enforced, 
 
\begin{equation}\label{eqn:gas_norm}
\frac{4}{3}\pi\rho_\mathrm{c}\left(r_\mathrm{c}y_\bullet\right)^3 \left._2F_1\left( \frac{3}{2},\frac{3}{2}\beta,\frac{5}{2};-y_\bullet^2 \right)  \right. = f_\mathrm{g}(m,z)~m~.
\end{equation} 
In the previous equation $y_\bullet\equiv R_\bullet/r_\mathrm{c}$, while $\left._2F_1(a,b,c;x)\right.$ is the Gauss hypergeometric function. This leaves the two free parameters $r_\mathrm{c}$ and $\beta$ only. For the latter I adopted the reference value $\beta = 2/3$, which is in good agreement with the average outer gas slopes in observed X-ray clusters \citep{NE99.1,OT04.1,VI06.1,CR08.1}. Some authors \citep{CR08.1} find a dependence of this outer slope on the gas temperature, in that hotter clusters would have a steeper slope as compared to cooler ones. However I did not include such a dependence, since the actual details of the gas distribution are likely to play a marginal role in the inner part of structures, which is of interest here. For the core radius $r_\mathrm{c}$ I refer to \citet*{MA98.1} (see also \citealt*{CA10.2}), where the authors have shown that a isothermal gas in hydrostatic equilibrium within a NFW dark matter halo is well represented by a $\beta$-model with a core radius of about one fifth of the scale radius. For a typical halo concentration this corresponds roughly to $r_\mathrm{c} = 0.05~R_\bullet$, which I adopted throughout. Please note that for my choice of $\beta = 2/3$, Eq. (\ref{eqn:gas_norm}) simplifies to

\begin{figure*}
	\includegraphics[width=0.43\hsize]{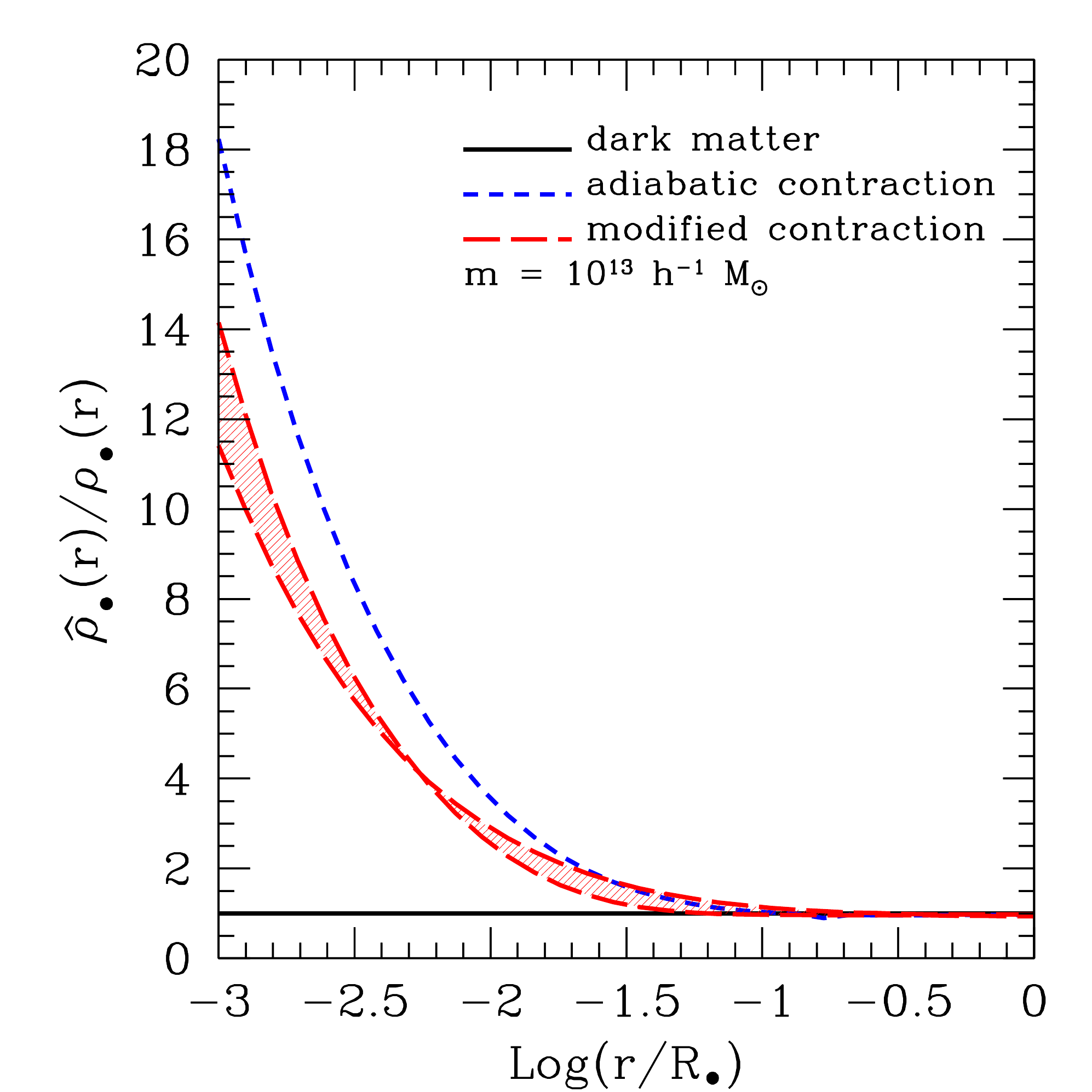}
	\includegraphics[width=0.43\hsize]{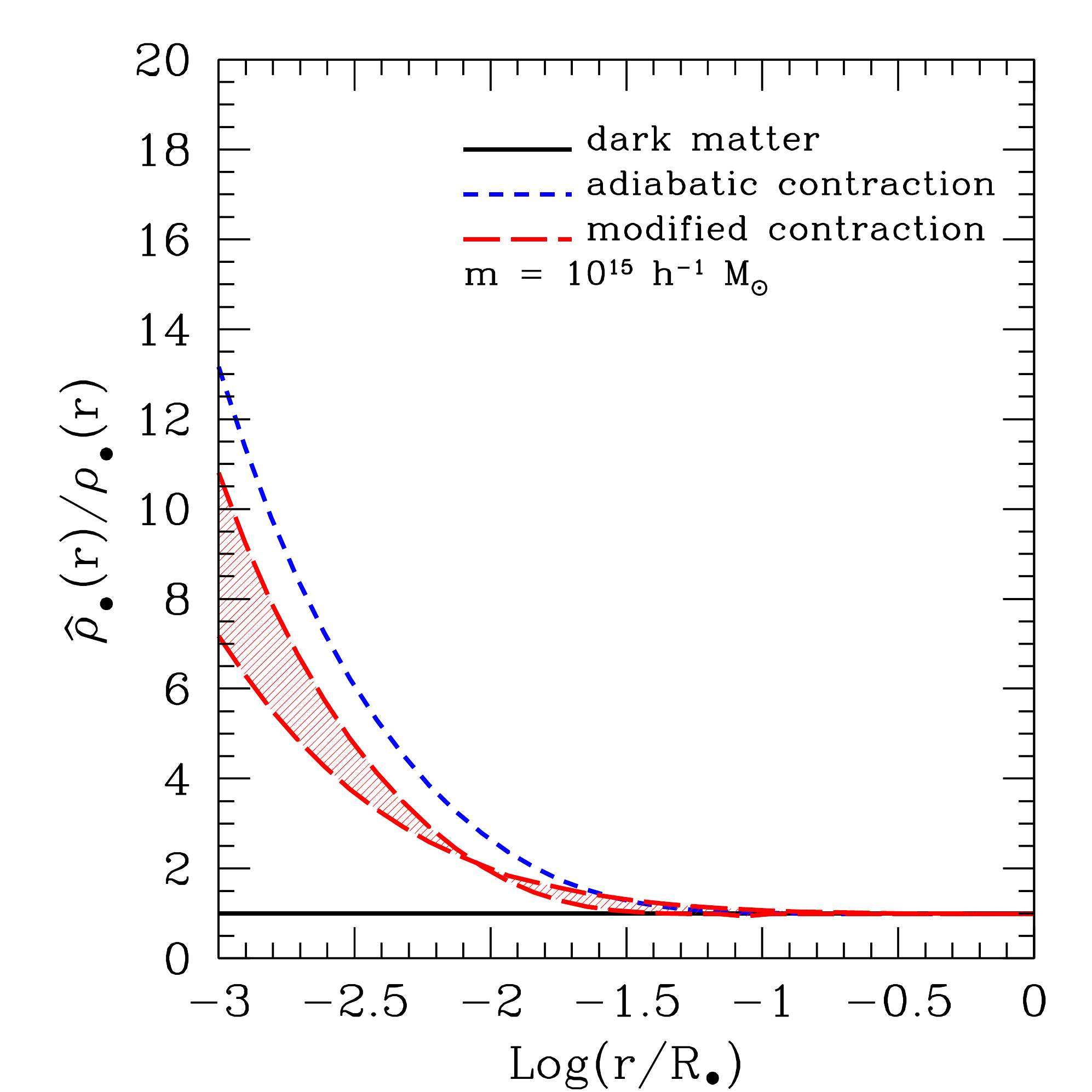}
	\caption{The contracted dark matter profiles divided by the original NFW profile, for a structure of total mass $m=10^{13}~h^{-1}M_\odot$ (left panel) and $m=10^{15}~h^{-1}M_\odot$ (right panel). The blue short-dashed line represents the standard adiabatic contraction, while the red shaded region enclosed by the two long-dashed lines shows the modified contraction model of \citet{GN11.1}. Note that the radial distance from the center of the halo is expressed in units of the virial radius \emph{of the original NFW halo}.}
\label{fig:baryonicContraction}
\end{figure*}

\begin{equation}
4\pi \rho_\mathrm{c}r_\mathrm{c}^3 \left[y_\bullet - \mathrm{arctg}~(y_\bullet) \right] = f_\mathrm{g}(m,z)~m~.
\end{equation} 
 
For the stellar component I assumed it to be radially distributed according to a \citet{JA83.1} profile, that reads
 
\begin{equation}
\rho_\star(\zeta) = \frac{\rho_0}{\zeta^2(1+\zeta)^2}~,
\end{equation}
where $\zeta\equiv r/r_0$. This equals at assuming all the stellar contribution to be concentrated in the Brightest Cluster Galaxy (BCG) of clusters and groups, ignoring the presence of other galaxies. This is a hypothesis that can certainly be lifted in future more detailed studies. For the moment I am interested only in keeping the model as simple as possible, and in providing a proof of concept regarding the impact of this stellar component on the concentration-mass relation. In Section \ref{sct:discussion} I elaborate more on this approximation and argue that a redistribution of stars at large radii is mostly degenerate with the assumed stellar mass fraction.

The \citet{JA83.1} profile has an isothermal slope of $-2$ at small radii, in agreement with the profile observed for stellar dominated systems such as elliptical galaxies \citep{KO09.2}. As before, the normalization of the profile is set by the stellar mass fraction, that is

\begin{equation}\label{eqn:star_norm}
4\pi\rho_0r_0^3\left( 1-\frac{1}{1+\zeta_\bullet} \right) = f_\star(m,z)~m~,
\end{equation}
where $\zeta_\bullet \equiv R_\bullet/r_0$. The remaining free parameter $r_0$, which equals the half-mass radius for the stellar distribution, should be a few percent of the virial radius of the dark matter halo. I adopt $r_0 = 0.02~R_\bullet$ henceforth, however results obtained by adopting different values for this free parameter will be shown. For a massive galaxy cluster this choice translates into $r_0 \simeq 20~h^{-1}$ kpc, which is a typical effective radius (\citealt*{SA02.1}; \citealt{SA05.1}). Because I assume that all stellar mass is included in this central clump, the value of $r_0$ could possibly be larger than this by up to $\sim 60\%$. I refer to Figure \ref{fig:concentrationMass_STELLAR_RADIUS} below to show that the impact of such a change is modest. Note that, since $r_0 \ll R_\bullet$, the left-hand side of Eq. (\ref{eqn:star_norm}) is actually very close to its asymptotic value $4\pi\rho_0r_0^3$.

In Figure \ref{fig:densityProfiles} I show the density profiles of the three matter components inside two systems of different masses, one typical of giant ellipticals/poor groups and the other characteristic of massive galaxy clusters. The different relevance of the various components can be clearly seen, with stars becoming dominant only at a few percent of the virial radius, while the hot diffuse gas remaining always subdominant with respect to the dark matter component. Also note how the gas component is much more important than the stellar component in the outskirts of structures, more so for more massive objects. It is significant that the stellar contribution starts having an influence at radii corresponding to the smallest ones probed by X-ray or gravitational lensing studies, because this implies that the apparently small differences seen between the total density profiles of structures with different masses are going to have a substantial impact. Turning the argument around, the choice of the smallest radius for profile fitting has a large relevance on the derived parameters.

\subsection{Baryonic contraction} \label{sct:contraction}

I mentioned above that the cooling of gas and the formation of large amounts of stars in the very central regions of cosmic structures is expected to drag along dark matter, thus effectively increasing the host halo concentration. There has been an intense debate in the literature about whether this contraction effect is real and to what extent it affects actual structures. While a variety of observational works find evidence for baryon contraction (\citealt{MI08.1,SC10.1,SO11.1}; see also additional references in \citealt{GN11.1}), several others recover cored dark matter profiles \citep{SA04.1,NE09.1,NE11.1}, so that the observational picture is not entirely clear at the moment. In the present paper I assumed that baryon contraction does take place. This is also supported by a very recent comprehensive study showing the occurrence and amplitude of the baryon contraction effect in a variety of simulations, presented by \citet{GN11.1}. Although this study is not the last word on the topic (see the discussion in Section \ref{sct:discussion} about the role of AGN feedback), it provides a convenient way to parametrize the impact of baryonic cooling as observed in numerical simulations.

Let me start by describing the standard adiabatic contraction model by \citet{BL86.1,RY87.1}, which assumes any structure as made of concentric spherical shells which contract isotropically by conserving angular momentum. Given this, the radius $r_1$ that initially encloses a dark matter mass $m_\bullet(r_1)$ can be related to the final radius $r_2$ enclosing the same mass by

\begin{figure*}
	\includegraphics[width=0.43\hsize]{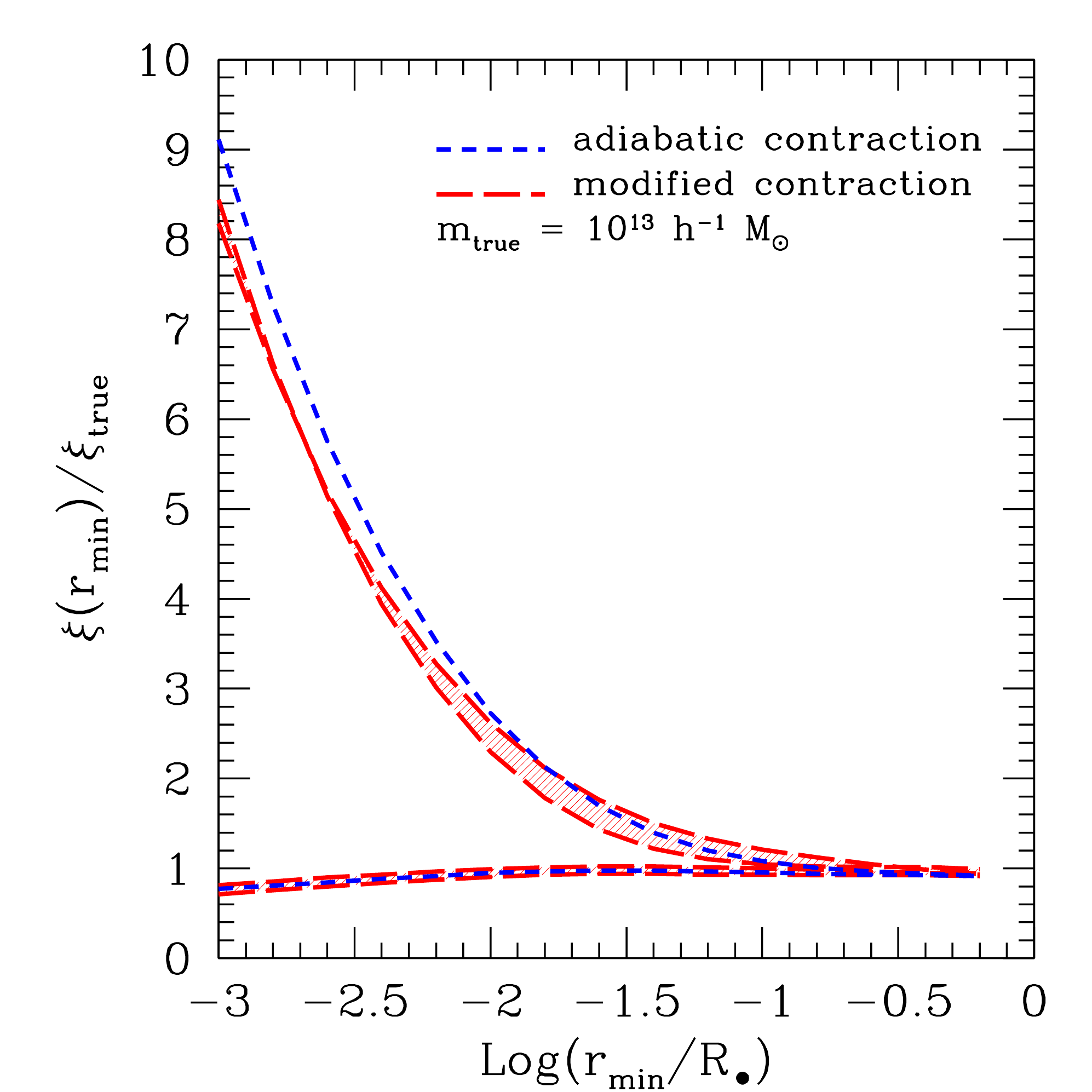}
	\includegraphics[width=0.43\hsize]{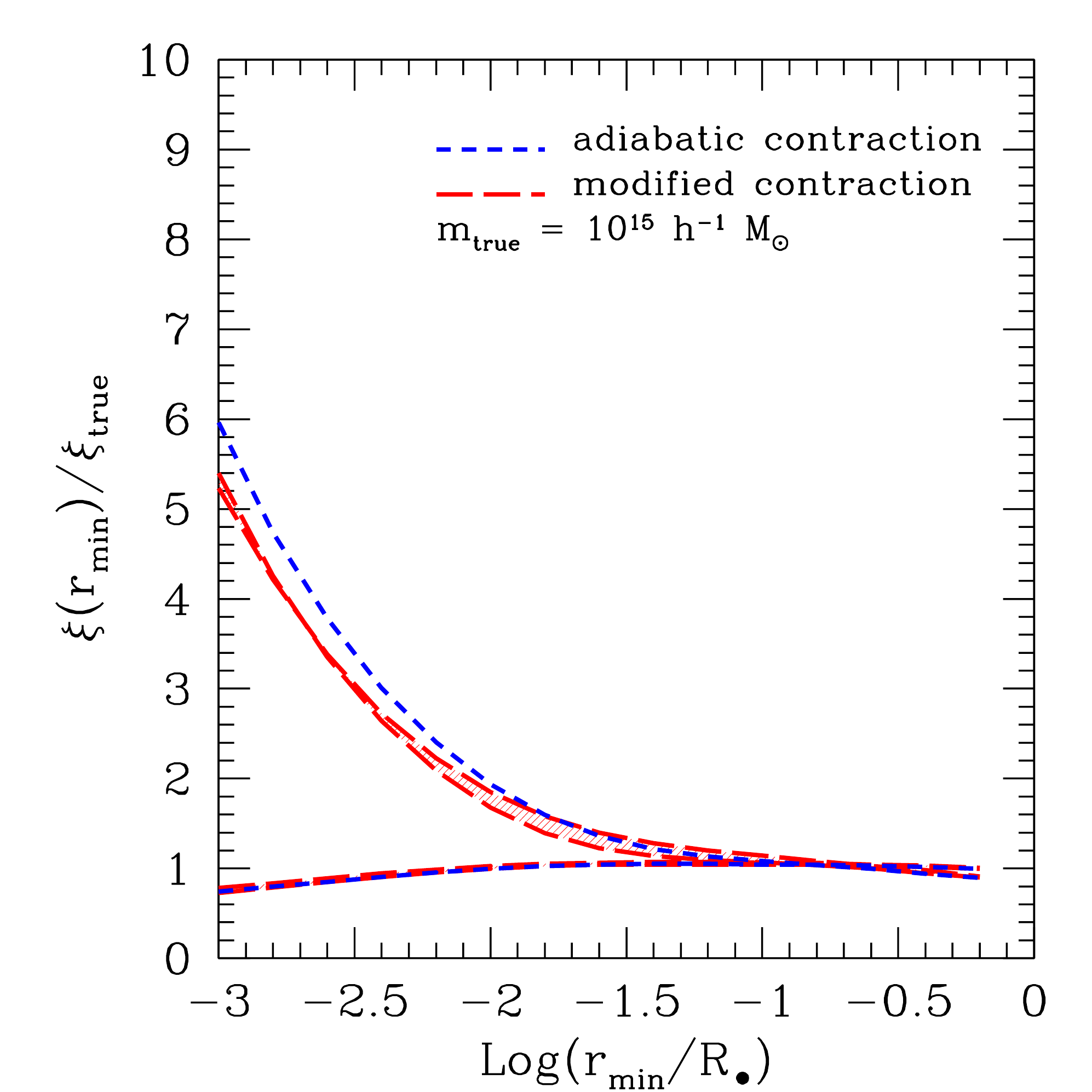}
	\caption{The masses and concentrations resulting from fitting an NFW profile to the total matter density profile, including baryonic contraction. The trends as a function of the minimum radius used for the fitting are shown. In both panels the lower set of curves refers to the mass ($\xi=m$), while the upper set represents the concentration ($\xi=c$) in units of their true values, that is the total mass of the structure and the concentration of a NFW profile with that mass according to \citet{GA08.1}. The blue short-dashed lines show the results of assuming the standard adiabatic contraction prescription, while the red shaded areas enclosed by the red long-dashed lines show the modified contraction model of \citet{GN11.1}. Finally, the left panel refers to $m_\mathrm{true} = 10^{13}~h^{-1}M_\odot$, while the right panel to $m_\mathrm{true} = 10^{15}~h^{-1}M_\odot$.}
\label{fig:nfwFit}
\end{figure*}

\begin{equation}\label{eqn:ad}
\left[m_\bullet(r_1) + m_{\mathrm{b},1}(r_1)\right]r_1 = \left[ m_\bullet(r_1)+m_{\mathrm{b},2}(r_2) \right]r_2~,
\end{equation}
where $m_{\mathrm{b},1}(r_1)$ is the total initial baryon mass within $r_1$, while $m_{\mathrm{b},2}(r_2)$ is the total final baryon mass within $r_2$. In order to determine the amount of dark matter contraction due to baryonic cooling it is hence necessary to establish initial and final mass profiles for baryons. I assumed that there are no stars at some initial time in the dark matter halos, so that all baryonic matter is composed by hot diffuse gas, $m_{\mathrm{b},1}(r_1) = m_{\mathrm{g},1}(r_1)$. The gas is still distributed as described in the previous Subsection \ref{sct:distribution}, only its profile is differently normalized by replacing $f_\mathrm{g}(m,z)$ in Eq. (\ref{eqn:gas_norm}) with $f_\mathrm{g}(m,z)+f_\star(m,z)$. After the cooling and star formation has occurred, stars and gas are distributed in the structure as described above, $m_{\mathrm{b},2}(r_2) = m_{\mathrm{g},2}(r_2)+m_{\star,2}(r_2)$.

As it has been shown by \citet{GN11.1}, the adiabatic contraction model always overestimates the actual amount of contraction measured in numerical simulations. Hence, the authors provided a simple modification to Eq. (\ref{eqn:ad}) that more accurately capture the contraction due to baryonic cooling (see also \citealt{GN04.1}). This modification reads

\begin{equation}
\left[m_\bullet(\bar r_1) + m_{\mathrm{b},1}(\bar r_1)\right]r_1 = \left[ m_\bullet(\bar r_1)+m_{\mathrm{b},2}(\bar r_2) \right]r_2~,
\end{equation}
where

\begin{equation}
\frac{\bar r}{0.03~R_\bullet} = A~\left( \frac{r}{0.03~R_\bullet} \right)^w~.
\end{equation}
\citet{GN11.1} suggested adopting $A = 1.6$, while they could not find a unique value of $w$ that would fit equally well the wide variety of simulations they were considering. This spread is probably ascribable to the variety of formation histories of cosmic structures, as well as to the different implementation of baryonic non-gravitational physics that different simulation works adopt. In what follows I shall consider a range of values for $w$ that covers the variety of results found by \citet{GN11.1}, that is $0.6 \le w \le 1.3$. Please note that if $A = 1$ and $w = 1$ the modified contraction scenario folds back to the standard adiabatic one.

\begin{figure*}
	\includegraphics[width=0.43\hsize]{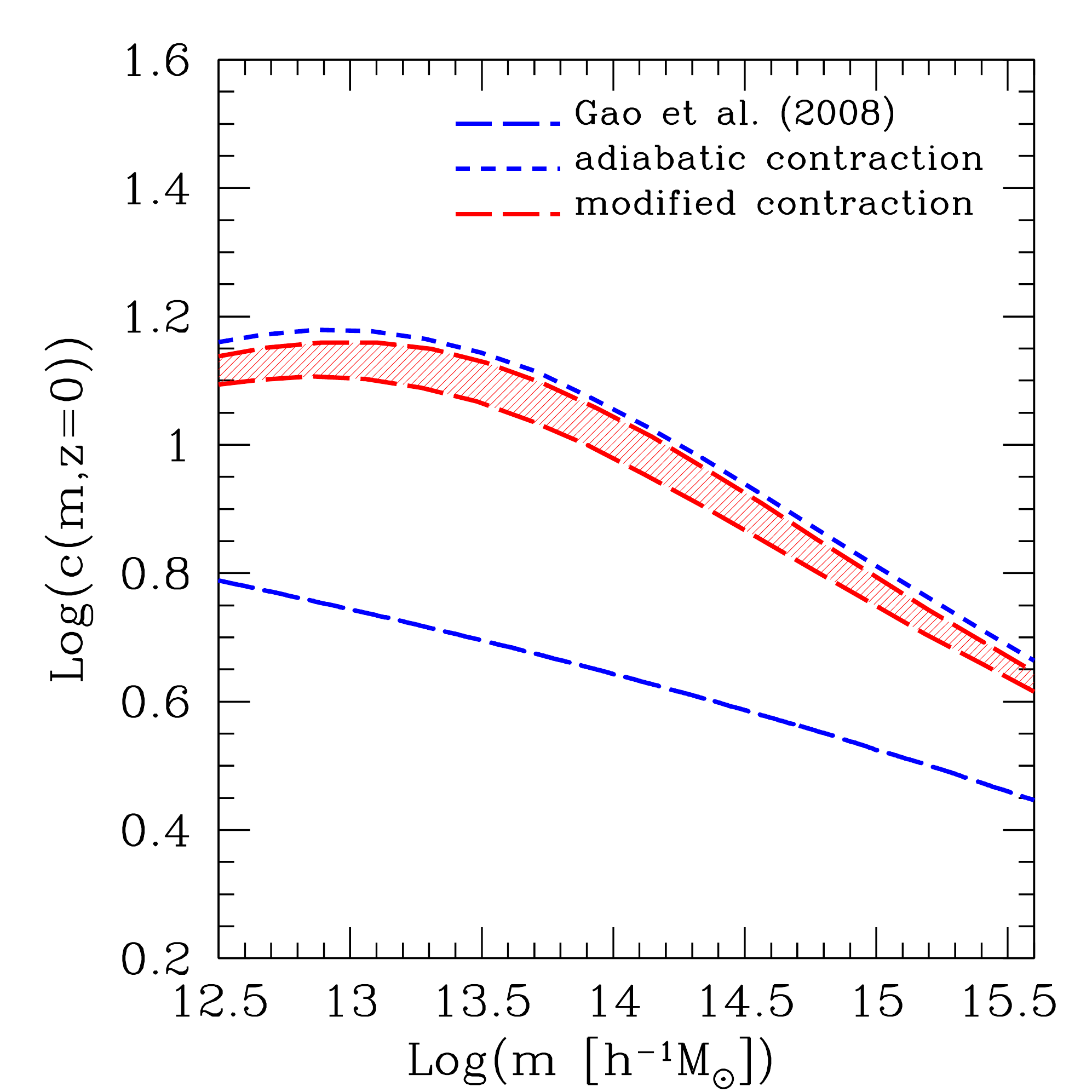}
	\includegraphics[width=0.43\hsize]{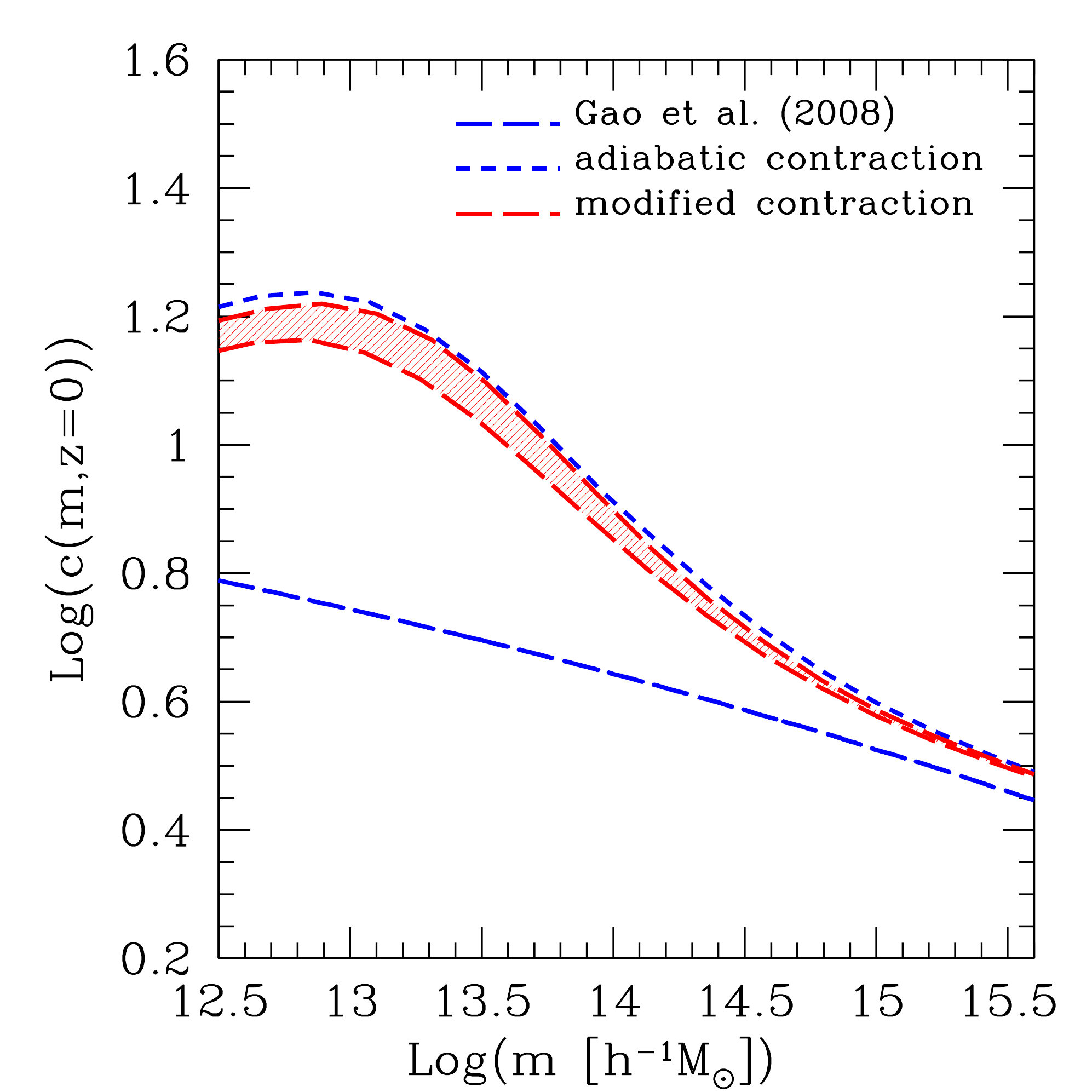}
	\caption{The concentration-mass relation resulting by including the effect of baryonic cooling and star formation. The blue short-dashed line represents the result of assuming the standard adiabtic contraction model, while the red shaded area bracketed by the red long-dashed lines refer to the modified contraction model of \citet{GN11.1}. For reference, the blue long-dashed curve shows the original relation found by \citet{GA08.1} by using $n$-body simulations. The left panel refers to the fiducial model for the baryon fraction as a function of mass, while the right panel considers a substantially reduced star formation in high-mass structures (see Figure \ref{fig:massFractions} and the related discussion in the text).}
\label{fig:concentrationMass}
\end{figure*}

In Figure \ref{fig:baryonicContraction} I show the effect of baryonic contraction on the dark matter profiles of structures with two different masses. As naively expected, the effect of contraction is milder for larger masses, due to the relatively smaller amount of baryonic cooling. Also, I observe that the parameter choice bracketing the actual contraction measured in various simulations by \citet{GN11.1} provides generically smaller dark matter accumulation in the central regions than the adiabatic contraction model, as indeed was found by the authors themselves. 

Although not visible in Figure \ref{fig:baryonicContraction}, the mass contraction actually becomes an expansion for the \citet{GN11.1} parameter choice $A=1.6$ and $w=1.3$ at $r\sim 0.3~R_\bullet$. The reason for this is that at this radius $\bar r \sim R_\bullet$, and the mass profiles are normalized so that initial baryonic mass and the final baryonic mass are equal at the virial radius of the original dark matter halo. Since however the final baryonic mass decreases more steeply than the initial one, outside the virial radius the former becomes smaller than the latter, thus effectively leading to a slight expansion. Finally, it should be noticed that the differences between the standard adiabatic contraction scenario and the modified baryonic contraction advocated by \citet{GN11.1} are more pronounced at very small radii, while they turn out to be relatively mild at $r \gtrsim 0.01~R_\bullet$, the radial range that shall be considered in what follows.

\section{Results}\label{sct:results}

I now turn to the main results of this work, namely describing how the mass and concentration of a galaxy system change with respect to the input values when fitting the total density profile, including the effect of baryonic contraction on dark matter. It is quite intuitive that the result of this procedure would depend on the radial range used in the fitting, with fits extending to smaller radii returning arguably larger concentrations, due to the stronger effects of star formation and baryonic contraction in the inner regions of structures. In order to address this issue, I computed the total density profiles, including baryonic contraction, for two objects having masses $m = 10^{13}~h^{-1}M_\odot$ and $m = 10^{15}~h^{-1}M_\odot$ at $n=16$ radii logarithmically spaced between $10^{-3}R_\bullet$ and $R_\bullet$. Then, I fitted NFW profiles to these density profiles, and compared the resulting masses and concentrations to their input values. I repeated this procedure $n-1$ times, each time removing the smallest radius from the fitting procedure. The results, as a function of the minimum radius adopted in the fitting are displayed in Figure \ref{fig:nfwFit}.

Let me focus first on the behavior of the mass. Introducing baryons and the effect of their cooling on a structure, while still fitting the total density profile with a NFW function causes the mass to be slightly underestimated. This effect is very mild however, being at most at the level of $\sim 20\%$ when the entire radial range $[10^{-3},1]~R_\bullet$ is considered. Additionally, I note that there is basically no difference between different prescriptions for the baryonic contraction. On the other hand, the behavior of the concentration is much stronger. Including very small radii in the fitting procedure results in substantially overestimated concentration values, up to almost an order of magnitude. As expected, this effect is largely reduced if only relatively large radii $r \gtrsim 0.05~R_\bullet$ are considered. On a related note, it is interesting to point out that in the observational study of \citet*{MA08.2}, who found a fair agreement with theoretical concentrations, a relatively large innermost radius has been adopted. I also observe that the concentration overestimation is larger for smaller masses and when the standard adiabatic contraction recipe is adopted, because in both cases the impact of baryonic cooling and star formation is larger than for, respectively, larger masses and the more realistic contraction prescription of \citet{GN11.1}. For $r\gtrsim 0.01~R_\bullet$, the value that will be used in the remainder of this work, the difference between different baryonic contraction recipes is relatively limited, being at most of a factor of $\sim 2-3$.

The results shown in Figure \ref{fig:nfwFit} are interesting in their own respect. In cases in which radial profiles can be measured over a wide range of radii, for instance via gravitational lensing, these results tell one what is the minimum radius one should consider in the fitting procedure in order to obtain meaningful values of the concentration. In cases where only a limited and relatively central radial range is accessible to observations, as for instance in X-ray studies, these results tell one how much the resulting concentration estimates are biased. Figure \ref{fig:nfwFit} also makes clear that in order to evaluate the impact of baryonic cooling on the concentration-mass relation, it is necessary to first specify a minimum radius for the fitting procedure, as different choices will most definitely lead to different results. In what follows I shall focus on the choice $r_\mathrm{min} = 0.01~R_\bullet$. For the most massive clusters this translate in $r_\mathrm{min}\sim$ a few tens of $h^{-1}$ kpc, comparable to, e.g., the innermost radius adopted in X-ray studies. This approach is obviously a simplification of the actual procedure, however I am interested here in deriving general conclusions. In more detailed future studies it will be easily possible to adapt the selected radial range to the observation one is interested to match.

I now turn to the main point of the present investigation, that is, understanding how the concentration-mass relation is affected by the presence and cooling of baryons. In Figure \ref{fig:concentrationMass} I show the relation between concentration and mass implied by the \citet{GA08.1} prescription, as well as its modification resulting from the impact of baryons, computed with different recipes for the contraction of dark matter halos. Also, in the two panels of the same Figure I present results for the fiducial baryonic fraction and its modified version (see Subsection \ref{sct:fraction}) that results in a substantially reduced stellar abundance in high-mass structure, both depicted in Figure \ref{fig:massFractions}. As expected from the previous analysis (see Figure \ref{fig:nfwFit}) the concentration at a fixed mass is always increased by the presence of baryons, more so for low mass objects. However, this trend is not monotonic, rather it reaches a broad maximum at around $m\sim 10^{13}~h^{-1}M_\odot$ and then decreases slowly for lower masses. The reason for this stands in the stellar mass fractions. As can be seen in Figure \ref{fig:massFractions}, the stellar fraction does reach a broad maximum at $m\sim 10^{13}~h^{-1}M_\odot$ as well. For masses smaller than that the relative abundance of stars decreases, and hence so does the impact of baryonic cooling on the concentration of structures.

Overall, the impact of baryonic cooling is of the expected form, resulting in both a steeper slope and higher normalization for the modified concentration-mass relation. However, when adopting the fiducial stellar mass fraction, the effect on the slope is not as pronounced as it would be suggested by the observed relations summarized in Section \ref{sct:observed}, especially those of \citet{ET10.1}, \citet{OG11.1}, and \citet{SC07.2}. When this is replaced with a stellar mass fraction that is steeper at the high mass end, as it seems to be suggested by certain studies (see the discussion in Subsection \ref{sct:fraction}), the concentration-mass relation gets steeper as well. Although the new relation is not well represented by a power law, by a qualitative comparison with Figure \ref{fig:concentration} it appears evident that the model having substantially less stars at large masses is in better agreement with observations. I shall make a more quantitative assessment in this respect further below.

\begin{figure}
	\includegraphics[width=\hsize]{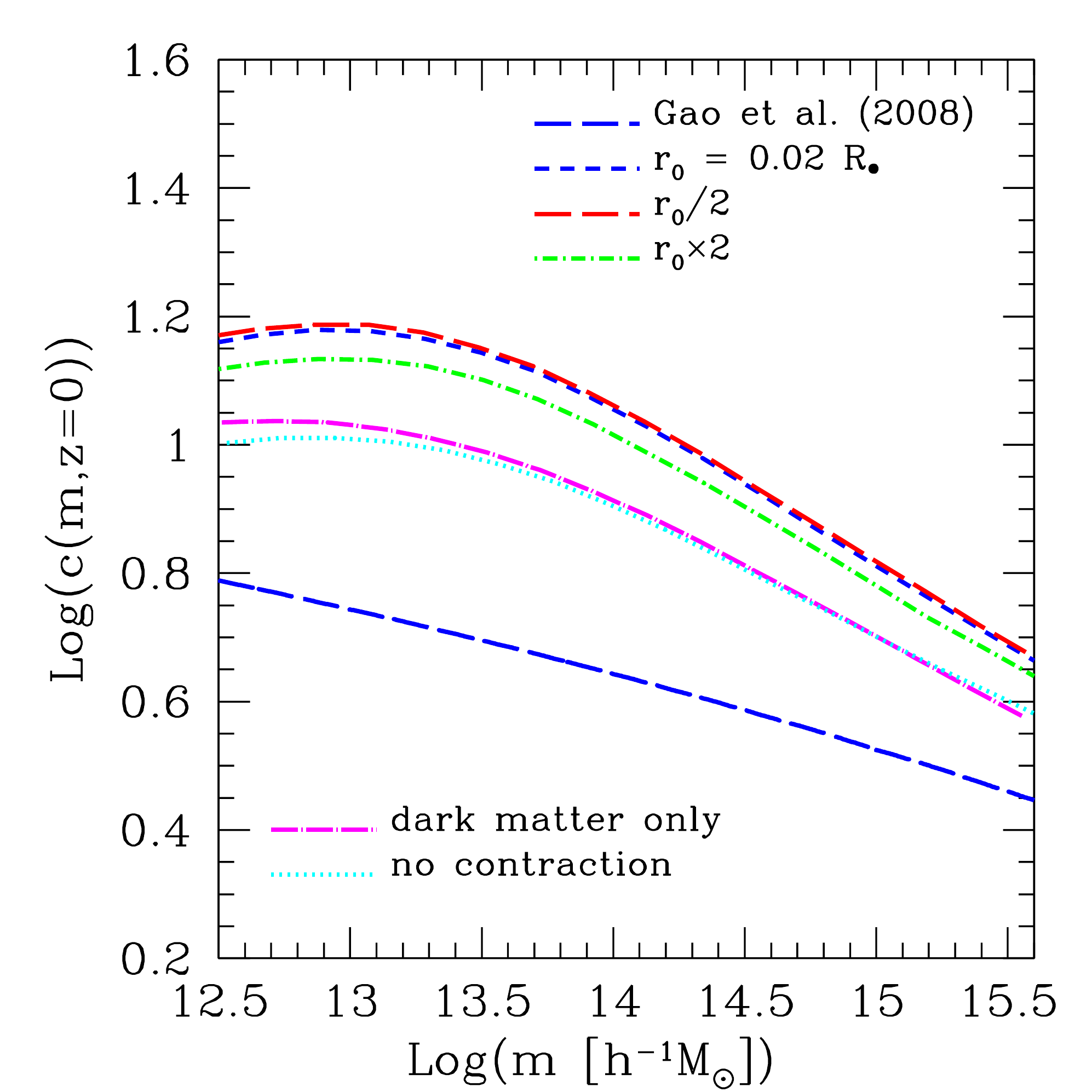}
	\caption{The concentration-mass relation according to the prescription of \citet{GA08.1} (blue long-dashed line) and according to the modified prescription presented here, adopting adiabatic contraction of dark matter halos. The blue short-dashed line refers to the fiducial value for the stellar scale radius, while the other lines show the results of modifications of the latter, as labeled in the plot. The magenta dot-long dashed line represents the relation obtained by fitting only the dark matter profile (after adiabatic contraction) with an NFW function, rather than the total mass profile. The cyan dotted line shows the result of fitting the total mass profile without any baryon contraction effect.}
\label{fig:concentrationMass_STELLAR_RADIUS}
\end{figure}

In Figure \ref{fig:concentrationMass_STELLAR_RADIUS} I show the impact of the scale radius $r_0$ of the stellar density profile on the modified concentration-mass relation. As can be easily understood, an increment in this scale radius results in a less compact stellar distribution. As a consequence the total mass profile is less peaked and the estimated concentration for a given mass tends to be smaller. The effect is however relatively limited, in that doubling $r_0$ results in a reduction of the best fit concentration of only $\sim 10\%$. One would expect the opposite effect when the stellar scale radius is decreased. However, while the overall mass profile does become more peaked, the effect of baryonic cooling also shifts at smaller radii, thus moving outside the radial range that is adopted for the NFW fitting. This latter effect counteracts the former, so that the resulting change in concentration is only very subtle. Given that changes due to the stellar distribution are comparable to or smaller than those induced by different prescriptions for the baryonic contraction of dark matter, in what follows I shall stick to the fiducial choice $r_0 = 0.02~R_\bullet$, and only show what happen if the baryonic contraction recipe is changed.

For completeness, in Figure \ref{fig:concentrationMass_STELLAR_RADIUS} I also report two additional lines. The magenta one shows the result of fitting an NFW function only to the dark matter distribution after contraction, rather than to the total mass distribution. The radial range is the same adopted above, the baryonic contraction is adiabatic, and the stellar mass fraction is the fiducial one. The resulting concentration-mass relation is rather similar in shape to the previous one, however it displays a substantially lower normalization. This is expected because the contribution of stars, which makes the overall density profile substantially peaked, is now ignored. The increment in concentration at a given mass is now driven only by the contraction of the dark matter density profile due to baryonic cooling. Analogous conclusions are reached if the modified stellar fraction is adopted instead. I note that in the simulations of \citet{DU10.1} baryonic physics causes the dark matter halo concentration to be overestimated by $\sim 10\%$ at most on cluster scales, while in Figure \ref{fig:concentrationMass_STELLAR_RADIUS} I find overestimates of $\sim 20-40\%$ on the same scales. This suggests either a baryonic contraction less efficient than suggested by the adiabatic model (as expected) or a stellar mass fraction significantly lower than predicted by the fiducial models adopted here (see below for further evidence in the same direction).  The cyan line in Figure \ref{fig:concentrationMass_STELLAR_RADIUS} displays what happen when fitting an NFW function to the total mass profile, but without any baryon contraction effect. Again, I adopted the same radial range used above and the fiducial stellar mass fraction. As can be seen, the inclusion of stars without any contraction of the dark matter profile and the inclusion of baryon contraction without any baryons have about the same effect on inferred concentrations at all masses, although the former results in a somewhat shallower relation. Note that these two effects cannot be simply summed together.

\begin{figure*}
	\includegraphics[width=0.43\hsize]{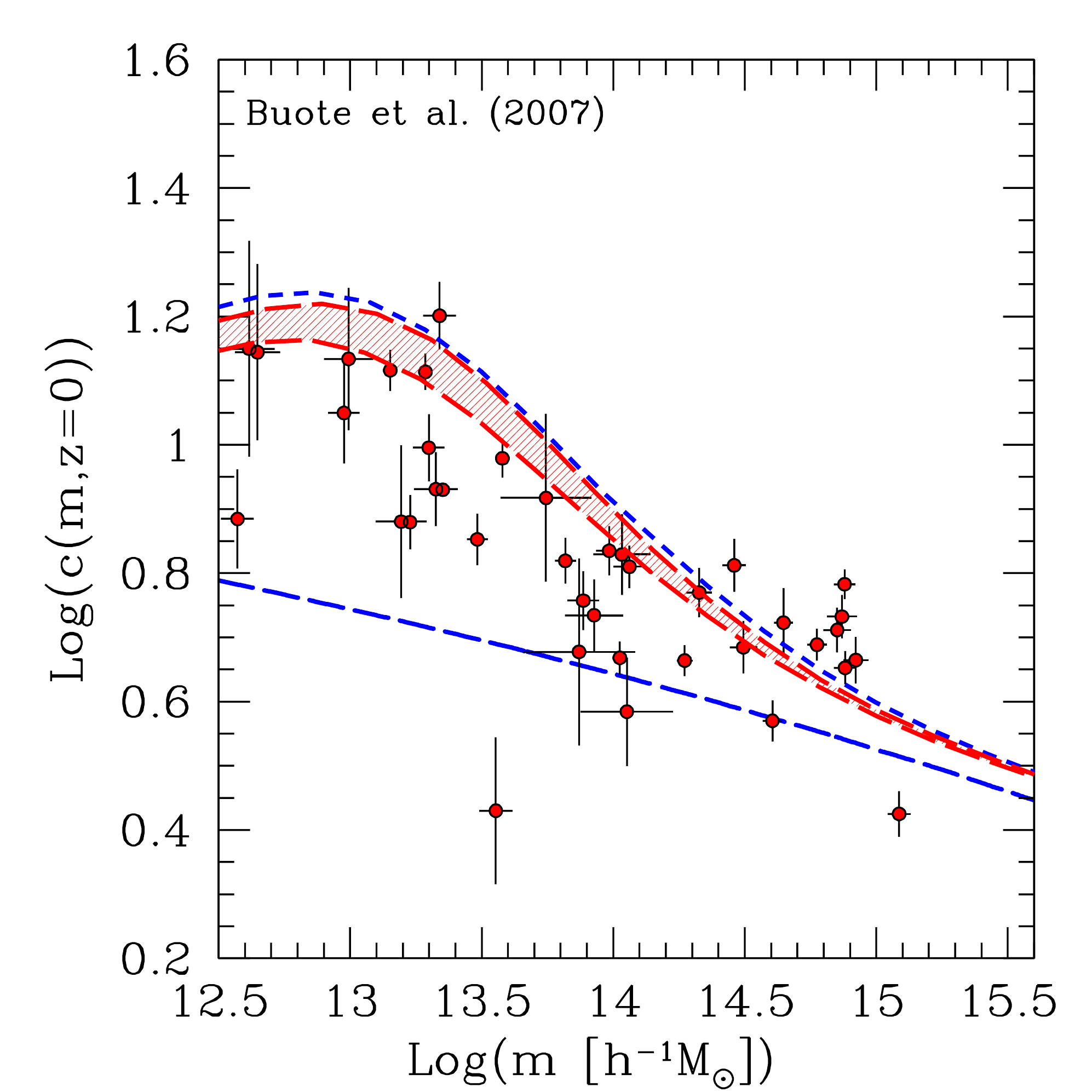}
	\includegraphics[width=0.43\hsize]{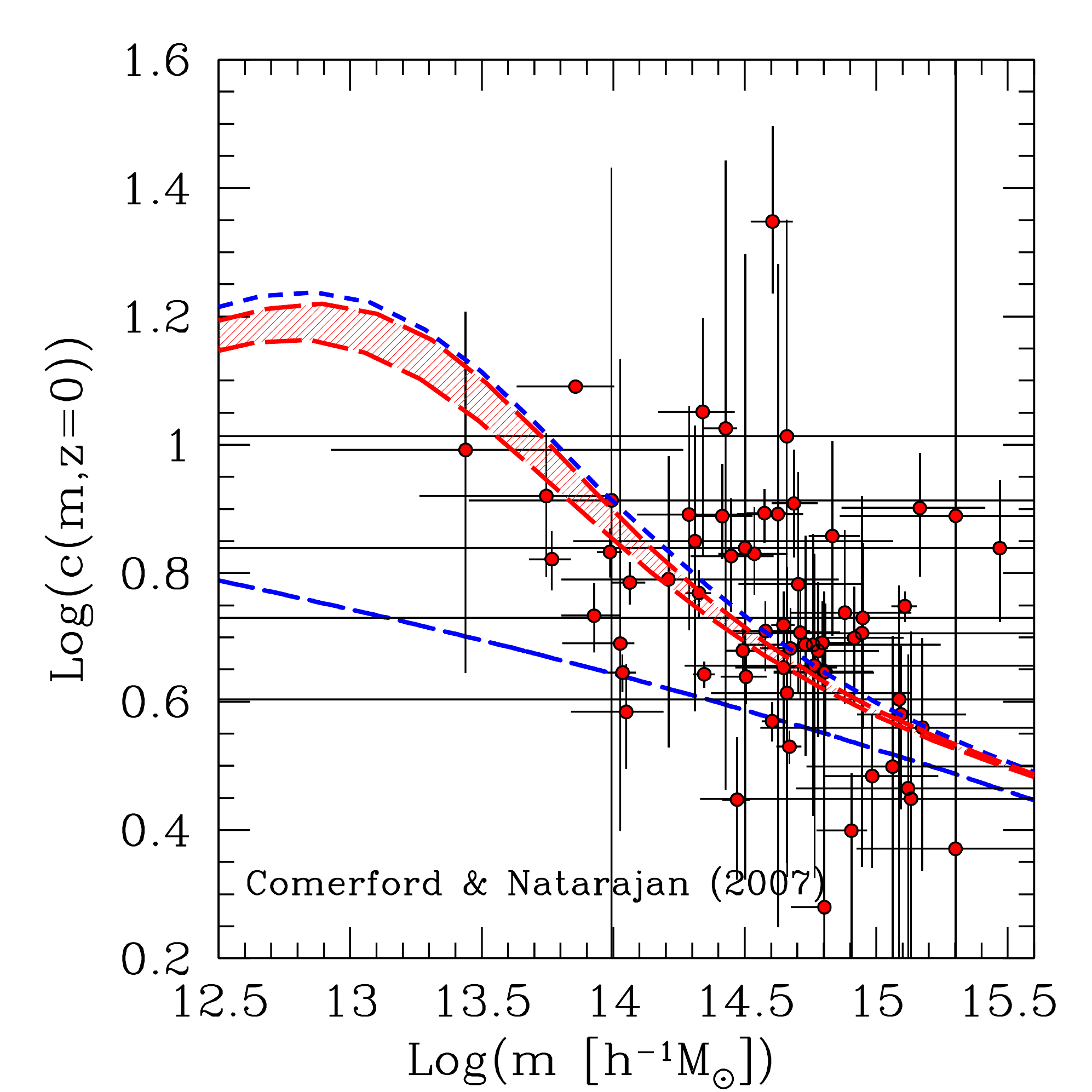}
	\includegraphics[width=0.43\hsize]{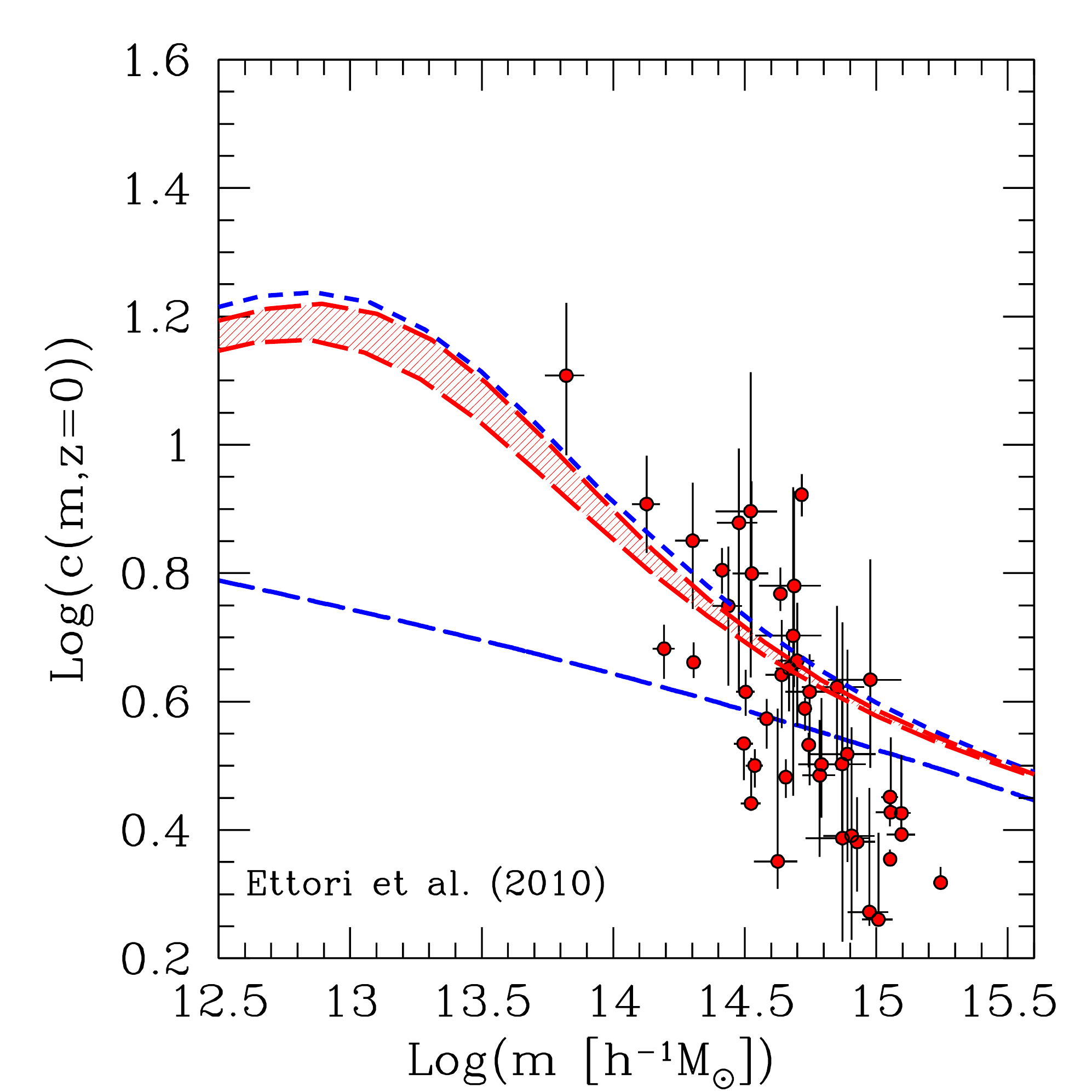}
	\includegraphics[width=0.43\hsize]{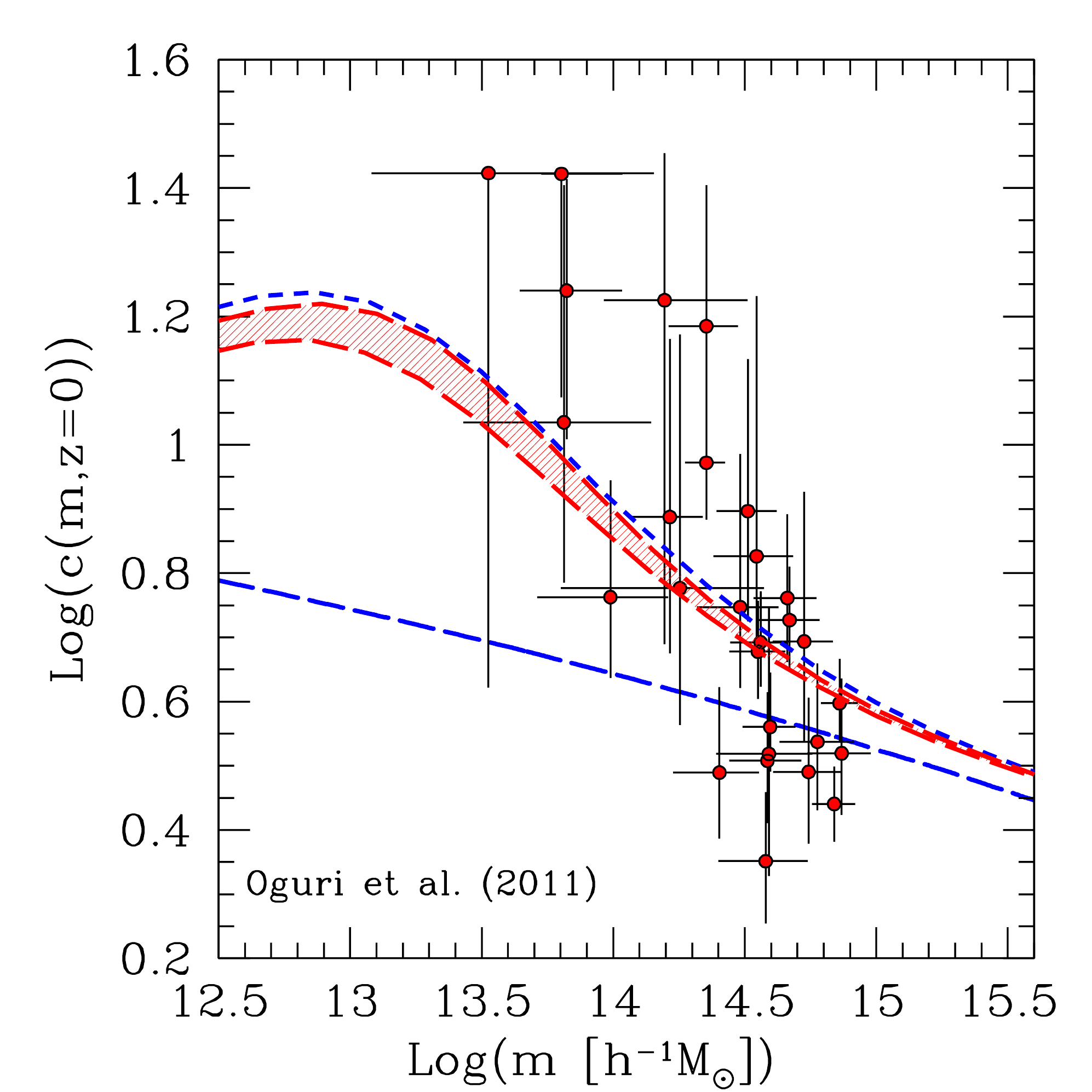}
	\includegraphics[width=0.43\hsize]{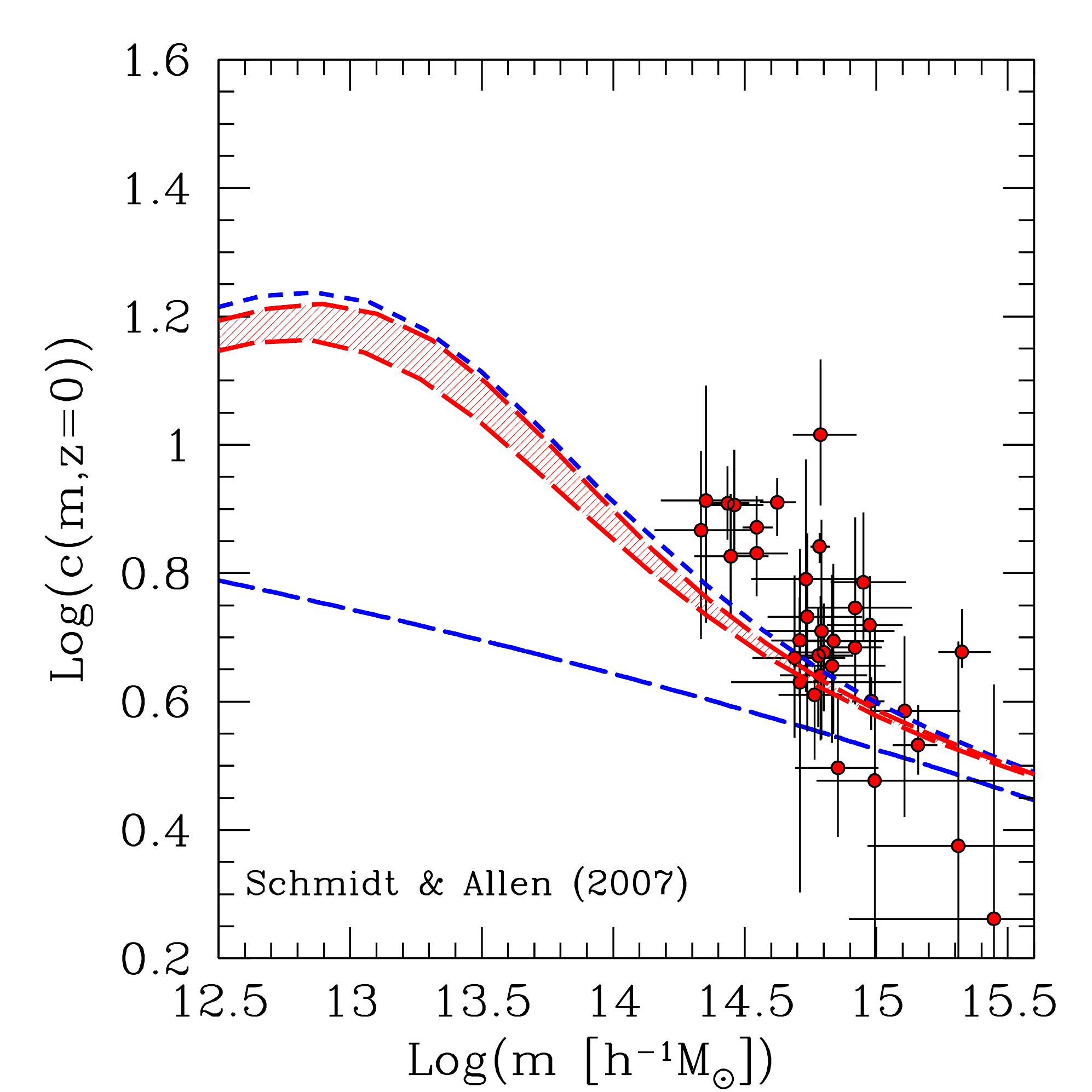}
	\includegraphics[width=0.43\hsize]{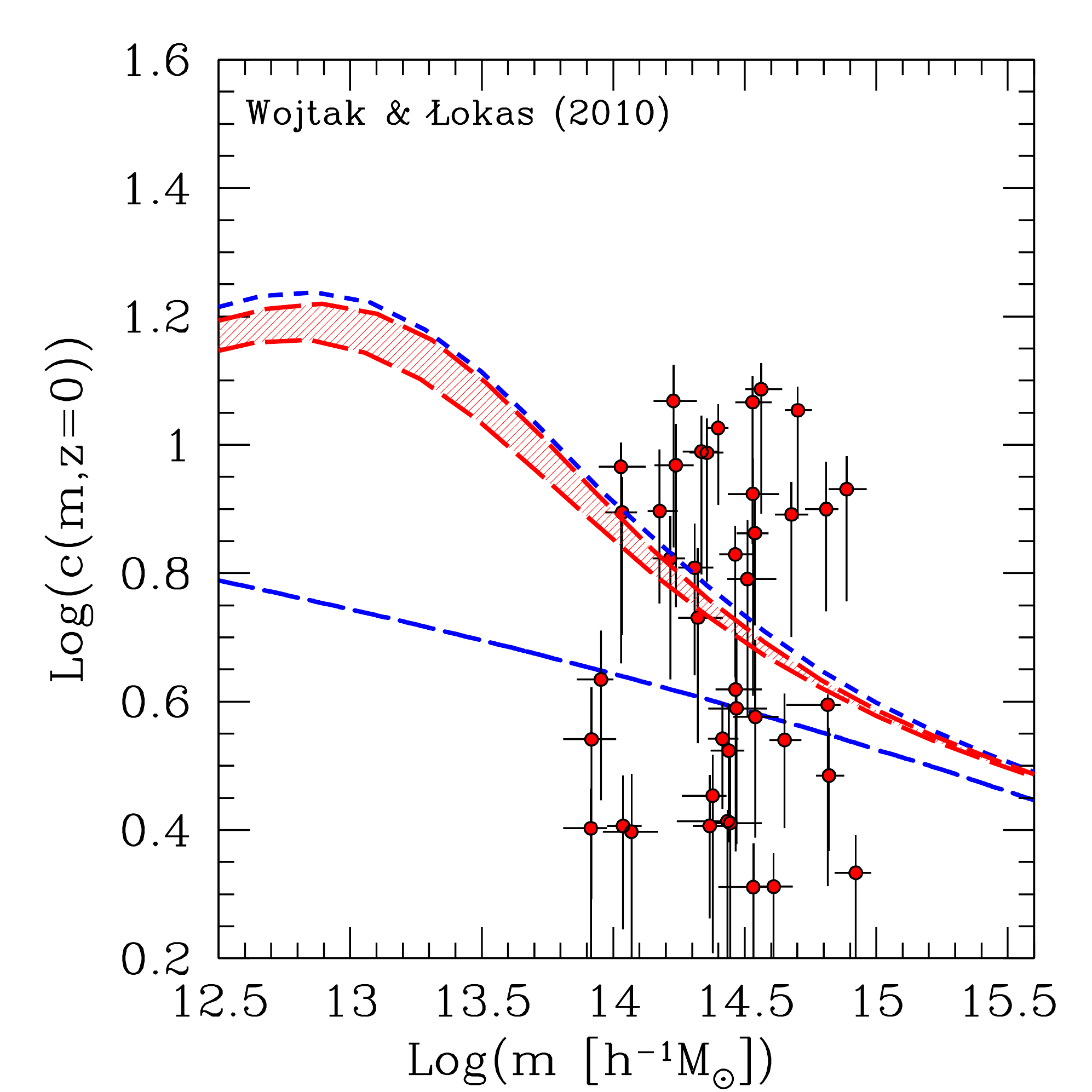}
	\caption{The observed relation between the concentration and the mass for groups and clusters of galaxies, as presented in Figure \ref{fig:concentration}. The blue long-dashed line represents the relationship predicted by Gao et al (2008) at $z = 0$. The other lines refer to the concentration-mass relation obtained by including the effect of baryonic cooling and star formation with different models for the contraction of dark matter (see the text and previous Figures for details).}
\label{fig:newConcentration}
\end{figure*}

In Figure \ref{fig:newConcentration} I show the same observed concentration-mass relations of Figure \ref{fig:concentration} compared with the modified theoretical prescription employed here. I show the results obtained with the reduced stellar fraction, as they provide a better agreement with observations (see below) than the fiducial one. As can be seen in the Figure, the modified concentration model does provide a better qualitative fit to the observed relations. Exceptions to this are given by the samples of \citet{BU07.1} and \citet{WO10.1}. In the former case the modified prescription seems to somewhat overestimate the observed concentrations at small masses, while in the latter case nothing conclusive can really be said, due to the large scatter in the data. 

In order to quantify if and at what level the new concentration-mass relation provides a better representation of the observed data, for each observational sample I computed the $\chi^2$ quantity defined in Eq. (\ref{eqn:chi2}), by replacing $\hat{c}(m,z=0~|~c_0,\alpha)$ first with the \citet{GA08.1} prescription, and then with the modified recipe adopted here. I used the adiabatic contraction model for reference, keeping in mind that suitably tuning the baryonic contraction design and/or the typical size of the stellar distribution (see Figure \ref{fig:concentrationMass_STELLAR_RADIUS}) can lead to better results. As it turns out the goodness of fit $g\equiv \sqrt{\chi^2}$ decreases in all cases, with the exception of the \citet{ET10.1} cluster catalog. This means that in almost all cases the modified concentration-mass relation including the impact of baryonic cooling is a better representation of observations as compared to pure $n$-body prescriptions. The improvement is only slight ($g = 8.9$ becomes $g = 8.6$) for the \citet{WO10.1} catalog, while in all other cases it is more substantial, the best improvement being obtained for the \citet{SC07.2} catalog (from $g = 6.7$ to $g = 3.9$). For the \citet{ET10.1} sample the goodness of fit is almost unchanged, going from $g=6.3$ to $g=6.5$, implying that the modified relation does not provide any better representation of that catalog. However, it should also be recalled that in this work the authors report the masses and concentrations of dark matter halos only, thus caution should be taken in the comparison. I get back to this issue in the next Section \ref{sct:discussion}.

If the modified concentration-mass relation used above is replaced by its version resulting from the fiducial baryon fraction, hence with more stars in high-mass structures, the situation is reversed. More precisely, the goodness of fit worsens in all cases except for the \citet{SC07.2} sample, for which it goes from the original $g = 6.7$ down to $g = 5.0$. This confirms the qualitative conclusion reached above, according to which a substantially reduced stellar fraction in high-mass structures constitutes a better fit to the data.

\section{Discussion}\label{sct:discussion}

As mentioned several times in the course of the paper, the work by \citet{ET10.1} reports the concentrations and masses of dark matter halos rather than whole structures, hence a comparison with the modified concentration-mass relation reported in Figure \ref{fig:newConcentration} is not entirely self-consistent. \citet{SC07.2} also reconstructed the dark matter density profiles separately, however in order to determine the concentration-mass relation they referred to the total mass distributions, hence making the previous discussion fair. As for the remaining pure X-ray study by \citet{BU07.1}, there the authors employ results from previous papers (e.g., \citealt{HU06.2,GA07.1}). However in these previous works the mass fitting was performed in several different ways, both including and excluding the stellar contribution, in the latter case effectively fitting the overall matter distribution. It is not clear which measurements were actually adopted in the study of \citet{BU07.1}, hence for the sake of completeness I put it on the same footing as the \citet{ET10.1} work for the purpose of  this Section.

By considering the modified concentration-mass relation resulting from fitting the dark matter distribution alone and again considering the modified stellar fraction that provides a suppressed stellar abundance at high masses, the \citet{ET10.1} goodness of fit improves slightly (from $g=6.3$ down to $g=5.7$), while the \citet{BU07.1} goodness of fit improves substantially, decreasing from $g=8.5$ to $g=5.0$. The reason for the latter improvement is clear. The modified concentration-mass relation presented in Figure \ref{fig:newConcentration} overestimates the observed data, while the original prescription based on $n$-body simulations underestimate them. On the other hand, by considering the dark matter component alone gives rise to a relation that is in between the two, and hence fits very well the data. Also in the \citet{ET10.1} case, the modified concentration-mass relation obtained by considering the dark matter distribution alone fits the observed data better than the original prescription and the modified one adopted in the previous Section. The improvement however is very modest: the observed relationship is simply too steep to obtain a good match with theoretical models.

Although introducing the effect of baryonic cooling and condensation does improve the agreement of theoretical models with observations, there is still substantial room for improvement. One detail in particular that seems difficult to reconcile are the concentration values at large masses, that at least in the X-ray studies of \citet{SC07.2} and \citet{ET10.1} (but also in the lensing study by \citealt{OG11.1}) seem to be significantly lower than expected even with respect to pure dark matter studies. 
On the modeling side, significant AGN activity could prevent the formation of a cuspy stellar profile, and at the same time would also somewhat flatten the dark matter distribution. The simulations of \citet{DU10.1} shown that this is indeed the case and, surprisingly, the effect seems to be slightly more pronounced for high-mass clusters rather than groups of galaxies, a trend that would go in the correct direction.

At the same time, dark matter halos extracted from $n$-body simulations also present a scatter in concentration at a fixed mass, which is around $\sim 0.15$ in logarithm or more, and which might bring the less extreme observed points into agreement with theory within the respective uncertainties. However the systematicty of the discrepancy might still be a concern. On the observational side it is well know that, even for the most relaxed clusters as those considered in the catalogs mentioned above, non-negligible subsonic gas bulk motions typically lead to an underestimate of the cluster mass based on the hydrostatic equilibrium hypothesis by $\sim 10-20\%$ (\citealt{AS03.1}; \citealt*{RA04.1}; \citealt{RA06.1,RA08.1,ME10.1}). This could contribute to steepen the observed concentration-mass relation, however only if this effect is not active on low mass objects, an instance that is not straightforwardly verified.

Overall, in the studies of \citet{SC07.2}, \citet{ET10.1}, and \citet{OG11.1}, the observed concentration-mass relation seems to be even steeper than what suggested by the correction due to star formation presented here. While for the latter work this additional discrepancy can be explained via a strong lensing bias, this does not apply for the X-ray-based studies. One possibility is that for these cluster samples the stellar mass fraction at masses $\sim 10^{14}~h^{-1}M_\odot$ is substantially larger than indicated by the simple model adopted here. However, this might be difficult to reconcile with the observations of the baryonic fraction summarized in Subsection \ref{sct:fraction}. The above mentioned underestimation of X-ray masses based on hydrostatic equilibrium also goes in the correct direction, however it is difficult to see why this should not be at work on the group scale. It should also be noted that \citet{ET10.1} excluded the central $50$ kpc of each cluster in their X-ray analysis. Since this is a larger fraction of the virial radius for low-mass objects than for high-mass ones, Figure \ref{fig:nfwFit} suggests that this additional selection effect would flatten the modified concentration-mass relation, thus going in the wrong direction.

One point that deserves to be discussed further is the impact of uncertainties on the distribution of gas and stars within dark matter halos. It is well known that gas profiles exhibit a wide range of densities in structure cores \citep{AR10.1}. Core densities lower than expected for a isothermal $\beta$-model would result in a reduction of the overall structure concentration. Still, the contribution of hot gas to the overall density profile is subdominant at all radii, and non-negligible only in the outer parts of the most massive clusters (see Figure \ref{fig:densityProfiles}). Therefore I do not expect this uncertainty to change any of the conclusions of this paper. For the stellar distribution the situation is different, since I assumed here that all stars are in a single clump at the center of the structure. Removing mass from this stellar clump and redistributing it as a NFW profile \citep*{LI04.1} could be more realistic. The net effect would be a decrement in the estimated concentration, more pronounced for systems that are less dominated by a Brightest Cluster Galaxy (BCG), that are massive clusters. This procedure and its effect are basically similar to the decrement in the stellar fraction at high masses described above, since satellite galaxies give a negligible mass contribution with respect to, and their distribution is shallower than dark matter \citep{BU12.1}. As a matter of fact we verified that a trend similar to the one depicted in the right panel of Figure \ref{fig:concentrationMass} could also be obtained via a redistribution of stars within the structure, rather than a change in the stellar mass fraction.

Another issue that might be relevant for the present discussion is the redshift dependence of the concentration for a fixed mass, that I have been ignoring throughout the paper. It is relatively well established in numerical simulations that the concentration for the high mass clusters is rather constant in time, implying that such a dependence cannot be advocated in order to explain the underconcentration that several studies indicate in that mass range. On the other hand such a redshift dependence, despite being substantially weaker than $\propto (1+z)^{-1}$, could flatten somewhat the observed relation when factored in. I verified this for the \citet{ET10.1} catalog, finding that this is not the case. The steepness of the observed relation is unchanged, and the goodness of fit does not change significantly.

Finally, it is useful to spend a few words on the redshift evolution of the modified concentration-mass relation that has been used here. Going at redshifts larger than zero, the average profiles of the various matter components would tend to change. For the dark matter this change is relatively well defined, being driven only by the evolution of the concentration. For other matter components this is less well established, for instance there is an ongoing discussion about the extent to which the stellar distribution is more compact at high-$z$ \citep*{TR11.1}. Another ingredient that might change with cosmic time is the baryon fraction, although no convincing evidence has been found for a significant evolution of both the gas and stellar mass fractions from $z\sim 1$ \citep{AL04.2,ET09.1,GI09.2}. In any case, once one's favorite recipe for the redshift evolution of these ingredients is set, it is straightforward to obtain a modified concentration-mass relation at any redshift.

\section{Summary and conclusions}\label{sct:conclusions}

In the present work I investigated the concentration-mass relation inferred from six different observed cluster catalogs, and tried to produce a modified theoretical relation that would take into account the role of baryonic cooling and star formation. The main results can be summarized as follows.

\begin{itemize}
\item In almost all cases the slope of the observed concentration-mass relation is substantially larger than theoretical predictions. The exceptions are given by the \citet{CO07.1} and \citet{WO10.1} catalogs. In the former the slope is compatible with theoretical expectations, but the normalization is significantly higher than for a WMAP-based standard cosmology. In the latter the concentration at a fixed mass displays a very large scatter.
\item The systematicty with which the slope/normalization of the observed concentration-mass relation exceeds the theoretical expectation indicates that low-mass clusters and groups of galaxies are over-concentrated, while massive systems have the expected concentration, or are possibly underconcentrated. Baryonic cooling affects more low-mass systems, hence it is expected to be a possible explanation for this discrepancy.
\item A simple model in which every structure is spherical and composed by stars, diffuse gas, and dark matter undergoing a contraction due to baryonic cooling does indeed result in a larger estimated concentration for a fixed mass. As expected the effect is more enhanced at the low-mass end, with concentration values being increased up to a factor of $2-3$ for light galaxy systems.
\item This simple model also straightforwardly permits to determine the bias introduced in the measurements of the mass and concentration of an object as a function of the radial range covered. This is going to be extremely useful for the correct interpretation of future observational results in light of theoretical expectations.
\item The overall behavior of the resulting new concentration-mass relation depends significantly on the details of the stellar mass fraction. Low stellar abundances (or, alternatively, more extended stellar distributions) in large systems seem to be favored because otherwise the concentrations of these systems would be substantially overestimated with respect to the observed values. This result can provide a contribution to sort out the ongoing controversy about the stellar mass fraction in galaxy clusters. 
\item The modified concentration-mass relation provides a better fit to the observed relation than theoretically motivated prescriptions in all cases (including those in which a steepening of the relation was not deemed necessary), with the exception of the sample studied by \citet{ET10.1}. 
\end{itemize}

Although very simplified in its nature, the modified concentration-mass relation that was proposed in this work provides a definite improvement over the standard prescriptions based on fits to $n$-body simulations. Since it straightforwardly allows to include one's favorite baryon fraction recipe or density profiles, it will prove extremely useful in the interpretation of future statistical studies on the matter distribution in galaxy groups and clusters. Possible advances include the introduction of a distribution of profiles for both stars and dark matter, as well as the role of ellipticity and substructures (satellite galaxies) on the fitting of a spherically averaged shape. As another example of additional effects that have not been included here, several authors (\citealt*{RU08.2}; \citealt{ZE11.1}) have shown that the presence of baryons causes a global rearrangement of the total matter density profile resulting in a somewhat increased concentration \emph{on top} of the standard baryonic contraction. On a related note, substantial and complementary effort should also come from numerical simulations implementing baryonic physics, of which one first example has been given by \citet{DU10.1}.

At the same time, it is important to enlarge the observed sample of clusters and groups with measured density profiles and reduced systematic uncertainties. The current situation shows a very large sample-to-sample variance, as well as substantial scatter within each individual sample. The study of large and uniformly selected cluster catalogs for which selection biases are well understood will be a key step toward a better understanding of the formation of structures in the Universe.

\section*{Acknowledgements}

I am grateful to M. Bartelmann, C. Giocoli, A. H. Gonzalez, A. Kravtsov, J. Merten, and L. Moscardini for reading the manuscript and providing useful comments that helped improving the presentation of this work. I also wish to acknowledge beneficial suggestions by two anonymous referees.

{\small
\bibliographystyle{aa}
\bibliography{./master}
}

\end{document}